\documentclass{aa}

\usepackage[utf8]{inputenc}
\usepackage{graphicx}
\usepackage[varg]{txfonts}

\bibpunct{(}{)}{;}{a}{}{,}

\begin{document}
\title{The inside-out quenching of the MHONGOOSE galaxy NGC 1371}

\author{S. Veronese\inst{\ref{astron},\ref{kapteyn}} \and W. J. G. de Blok\inst{\ref{astron},\ref{kapteyn},\ref{uct}} \and F. Fraternali\inst{\ref{kapteyn}} \and F. M. Maccagni\inst{\ref{inafca}} \and J. Healy\inst{\ref{manchester},\ref{uksrc}} \and D. Kleiner\inst{\ref{astron}} \and T. A. Oosterloo\inst{\ref{astron},\ref{kapteyn}} \and R. Morganti\inst{\ref{astron},\ref{kapteyn}}}

\institute{Netherlands Institute for Radio Astronomy (ASTRON), Oude Hoogeveensedijk 4, 7991 PD Dwingeloo, The Netherlands, \email{veronese@astron.nl}\label{astron}
\and Kapteyn Astronomical Institute, University of Groningen, PO Box 800, 9700 AV Groningen, The Netherlands\label{kapteyn}
\and Department of Astronomy, University of Cape Town, Private Bag X3, 7701 Rondebosch, South Africa\label{uct}
\and INAF – Osservatorio Astronomico di Cagliari, Via della Scienza 5, 09047, Selargius, CA, Italy\label{inafca}
\and Jodrell Bank Centre for Astrophysics, School of Physics and Astronomy, University of Manchester, Oxford Road, Manchester M13 9PL, UK\label{manchester}
\and United Kingdom SKA Regional Centre (UKSRC), UK\label{uksrc}}

\date{Received <date> / Accepted <date>}

\abstract{We present the deepest 21-cm spectral line and 1.4 GHz broad-band continuum observations of nearby early-type spiral galaxy NGC 1371 as part of the MeerKAT H{\sc i} Observations of Nearby Galactic Objects: Observing Southern Emitters (MHONGOOSE) survey. We found the neutral atomic hydrogen (H{\sc i}) mostly distributed in a regularly rotating disc with a hole $\sim5$ kpc wide around the galactic centre. The continuum observations reveal, within the H{\sc i} hole, emission from one of the lowest luminosity AGN known to date and from two unique $\sim10$-kpc wide bipolar bubbles never observed before in this galaxy. The properties of the bubbles suggest that they may result from the impact of the low-power radio jet propagating within the gaseous disk instead of perpendicular to it. We found indication for jet-induced ionised outflows within the H{\sc i} hole but no molecular gas (upper limit of $M_{\text{H$_2$}}<2\times10^5\text{ M$_\odot$}$) is detected. The emerging picture is that the gas in the central regions has been rapidly depleted by the stellar bar or, despite its low power, the AGN in NGC 1371 is efficiently heating and/or removing the gas through the jets and possibly by radiative winds, leading to the inside-out quenching of the galaxy.}

\keywords{Galaxies: evolution - Galaxies: formation - Radio lines: galaxies - Line: identification}
\maketitle

\section{Introduction}\label{sec:intro}
There is a general consensus that galaxy formation and evolution are governed by two major physical processes: gas accretion (reviews by \citealt{putman12,tumlinson17,peroux20,tacconi20,faucher23}) and feedback (reviews by \citealt{fabian12,heckman14,harrison18,tacconi20}). Indeed, the time scale for the depletion of gas in galaxies is less than a Hubble time \citep{saintonge13,tacconi13,saintonge16,saintonge22}, which requires gas accretion to explain the measured cosmic star formation history \citep{fraternali12,madau14,dave17,walter20,chen21}, while feedback is needed to regulate the growth of galaxies \citep{silk12,sommerville15}. It is not known in detail how these two mechanisms interact throughout the lifetime of the Universe to assemble galaxies and produce the diversity of structures that we observe at the present age. 
\\\indent In our current understanding of gas accretion, pristine gas from the cosmic web or from the galaxy halo and mergers supplies the fuel for star formation over cosmic time. How this gas accretion occurs is an open question from both a theoretical and an observational perspective. One expectation is that for low-mass galaxies (halo mass $<5\times10^{11}$ M$_\odot$) it happens in a `cold mode' \citep{keres05,keres09,dekel06,dekel09,voort11,danovich15,nelson16}, where clouds and filamentary streams of cold ($T<10^5$ K) gas funnel onto galaxies and fuel the star formation. Conversely, high-mass galaxies are expected to be fed by `hot-mode' accretion, in which the gas falling onto galaxies forms a hot Circum Galactic Medium (CGM). This happens when the cooling time of the gas is larger than the free-fall time. When a hot CGM is formed, it can further cool through cooling flows \citep{thompson16,schneider18,schneider20,fielding22}, thermal instabilities \citep{mccourt12,sharma12,gaspari18} and induced cooling \citep{fraternali06,fraternali17,marinacci10,marinacci11,li21,marasco22}. It is unclear whether most of the accretion occurs in a cold or hot phase, and, observationally, direct unambiguous evidence of gas accretion, both for the hot and cold modes, is still missing \citep{braun04,wolfe13,wolfe16,das20,das24,kamphuis22,xu22,liu23}.
\\\indent If our understanding of gas accretion is limited, this also applies to our knowledge of feedback. Feedback comes in two forms: stellar \citep{tacconi20} and Active Galactic Nuclei (AGN; \citealt{fabian12,heckman14,harrison18}). Stellar feedback acts as a regulator of star formation \citep{schaye04,krumholz09,ostriker10,dave12,lilly13,leroy13,peng14,sommerville15,hayward17,tabatabaei18}. When star formation is high, supernova explosions inject energy into the galaxy, heating the interstellar medium (ISM) and expel gas from the galaxy via winds. This leads to the suppression of future star-forming activities. As star formation terminates, gas falls back onto the galaxy, cools down and collapses, restarting the stellar production. However, many of the details of this process are still uncertain and recent observational \citep{samal14,egorov17} and theoretical \citep{yu20,egorov23} studies proposed that supernova feedback may also be positive, locally enhancing the star formation rate.
\\\indent The role of AGN feedback is also uncertain. It is invoked in cosmological simulation to quench star formation and supermassive black hole (SMBH) growth to match the predicted properties of massive galaxies \citep{eagle,khandai15,dubois16,mccarthy17,nelson18,simba}. From a physical point of view, this negative feedback arises from the strong winds and jets launched by the central engine \citep{dimatteo05,springel05,bower06,croton06,ciotti10,wagner12,pontzen17,zinger20}. These outflows blow out and heat the gas, preventing its cooling and collapse and suppressing star formation. However, some theoretical studies propose that AGN feedback may also have a positive effect, promoting star formation \citep{ishibashi12,zinn13,silk13,muk18}. This happens because the jet winds compress and shock-heat the gas, which allows for very efficient radiative cooling. Empirically, both negative and positive AGN feedback have been observed in a number of different galaxies (see, e.g., \citealt{marasco20,venturi21,konda23,heckman23,heckman24} for negative feedback and \citealt{santoro15a,cresci15a,cresci15b,gallagher19,zhuang20} for AGN-induced star formation), making our understanding of AGN feedback more ambiguous.
\\\indent Both stellar and AGN feedback effects may be constrained by observational neutral atomic hydrogen (H{\sc i}) studies (see \citealt{tamburro09,ostriker10,ianjam12,stilp13,kim22} for stellar feedback, \citealt{morganti03,morganti16,morganti18,gupta06,chandola11,mahony13,gereb15a,maccagni17} for AGN feedback), yet these are limited by the need for both column density sensitivity and angular and spectral resolution. The recent MeerKAT H{\sc i} Observations of Nearby Galactic Objects: Observing Southern Emitters (MHONGOOSE, \citealt{mhongoose2}) survey represents a major step forward in understanding gas accretion and feedback in the context of galaxy formation and evolution. Using the high column density sensitivity (down to a few times $10^{17}$ cm$^{-2}$), angular resolution (up to $7''$), spectral resolution (1.4 km s$^{-1}$) and wide-field coverage (1.5$^\circ$) of the MeerKAT radio telescope\footnote{The MeerKAT telescope is operated by the South African Radio Astronomy Observatory, which is a facility of the National Research Foundation, an agency of the Department of Science and Innovation.} \citep{meerkat}, MHONGOOSE studies the distribution and dynamics of H{\sc i} in and around 30 nearby star-forming galaxies. We refer to \citet{mhongoose2} for information on MHONGOOSE and its selection criteria.
\\\indent In this paper, we will discuss the MHONGOOSE galaxy NGC 1371, a nearby early-type spiral galaxy member of the Eridanus group \citep{brough06,for23} (see Table \ref{table:obj} for the main observational properties). With a stellar mass of $4.3\times10^{10}$ M$_\odot$ and a combined mid-IR and UV Star Formation Rate (SFR) of $0.24$ M$_\odot$ yr$^{-1}$ \citep{cluver25}, it is the only target in the MHONGOOSE sample that is significantly below the main sequence of galaxy star formation \citep{mhongoose2},with a SFR that is $\sim1.7$ M$_\odot$ yr$^{-1}$ lower than expected. This statement holds even when considering different SFR estimates \citep{grundy23}. Photometric measurements of the $\lambda\sim154$ nm (FUV) and $\lambda\sim623$ nm (r) emission from the Galaxy Evolution Explorer (GALEX; \citealt{galex}) and the Dark Energy Survey (DES; \citealt{des}) place this galaxy at the edge between the star-forming main sequence and the green valley.
\\\indent It is unclear whether NGC 1371 hosts an AGN. Based on the H$\alpha$ emission, \citet{hameed99} claimed that the nuclear ISM is probably heated by the ionising UV radiation from post-asymptotic giant branch stars rather than by AGN activity. The lack of nuclear activity was corroborated by \citet{cisternas13}, who analysed the central X-ray emission with the Chandra X-ray Observatory \citep{chandra}. However, based on the same Chandra observations, \citet{hughes07} asserts that the H$\alpha$-X-ray luminosity correlation in the NGC 1371 nucleus is compatible with the relation found for the AGN by \citet{ho01}, suggesting that this galaxy hosts a low-luminosity AGN. Studying the $\sim1.4$ GHz continuum with the Giant Metrewave Radio Telescope (GMRT; \citealt{gmrt}) \citet{omar05} also claimed that NGC 1371 hosts an AGN. Evidence for this is a radio excess with respect to the far infrared emission \citep{yun01}, the observation of a bright radio source coincident with the nuclear H$\alpha$ emission, and a kpc-scale jet-like structure with no H$\alpha$ correspondence, suggesting that the extended radio emission does not come from star-forming regions. These findings are corroborated by \citet{grundy23} using the more sensitive and higher resolution 1.4 GHz continuum observation from the Widefield ASKAP L-band Legacy All-sky Blind surveY (WALLABY; \citealt{wallaby}). As such, NGC 1371 is one of the most tantalizing MHONGOOSE galaxies to study the effect of stellar and AGN feedback and possible quenching mechanisms. For example, environmental effects such as tidal interaction and gas stripping (see \citealt{hirschmann14,hatfield17} for theoretical works and \citealt{peng15,knobel15} for observational evidence), as well as secular processes such as bar quenching \citep{khoperskov18,george19} or disk instabilities \citep{bluck22} may suppress its star formation.
\\\indent This paper is divided as follows: in Sect. \ref{sec:data} we present the MHONGOOSE data used in this study; in Sect. \ref{sec:ana} we describe how these data were analysed to derive the main H{\sc i} and radio properties of this galaxy; in Sect. \ref{sec:disc} we discuss the results, focussing on the feedback mechanism that led to the suppression of star formation; in Sect. \ref{sec:conc} we summarised the conclusions and briefly provide some future prospects.

\begin{table}
    \caption{NGC 1371 main observational properties.}
    \label{table:obj}
    \centering
    \begin{tabular}{c c c}
    \hline\hline
    Parameter & Value & References \\
    \hline
    R.A. (J2000)          & 03$^{h}$ 35$^{m}$ 01.1$^{s}$  & (1) \\
    DEC. (J2000)          & -24$^{d}$ 55$^{m}$ 59.6$^{s}$ & (1) \\
    $v_\text{sys}$        & 1456.4 km s$^{-1}$            & (2) \\
    $D$                   & 22.7 Mpc                      & (3) \\
    $M_{\text{H{\sc i}}}$ & $9.5\times10^{9}$ M$_\odot$   & This work\\
    $M_\star$             & $4.3\times10^{10}$ M$_\odot$  & (4)\\
    SFR                   & 0.24 M$_\odot$ yr$^{-1}$      & (4)\\
    m$_{\rm FUV}$         & 14.4 mag                     & (5)\\
    m$_{\rm r}$           & 10.9 mag                    & (6)\\
    \hline
    \end{tabular}
    \tablebib{(1) \citet{evans10}; (2) \citet{mhongoose2}; (3) \citet{leroy19}; (4) \citet{cluver25}; (5) \citet{abbott21}; (6) \citet{bouquin18}.}
\end{table}

\begin{figure*}
    \centering
    \includegraphics[width=\hsize]{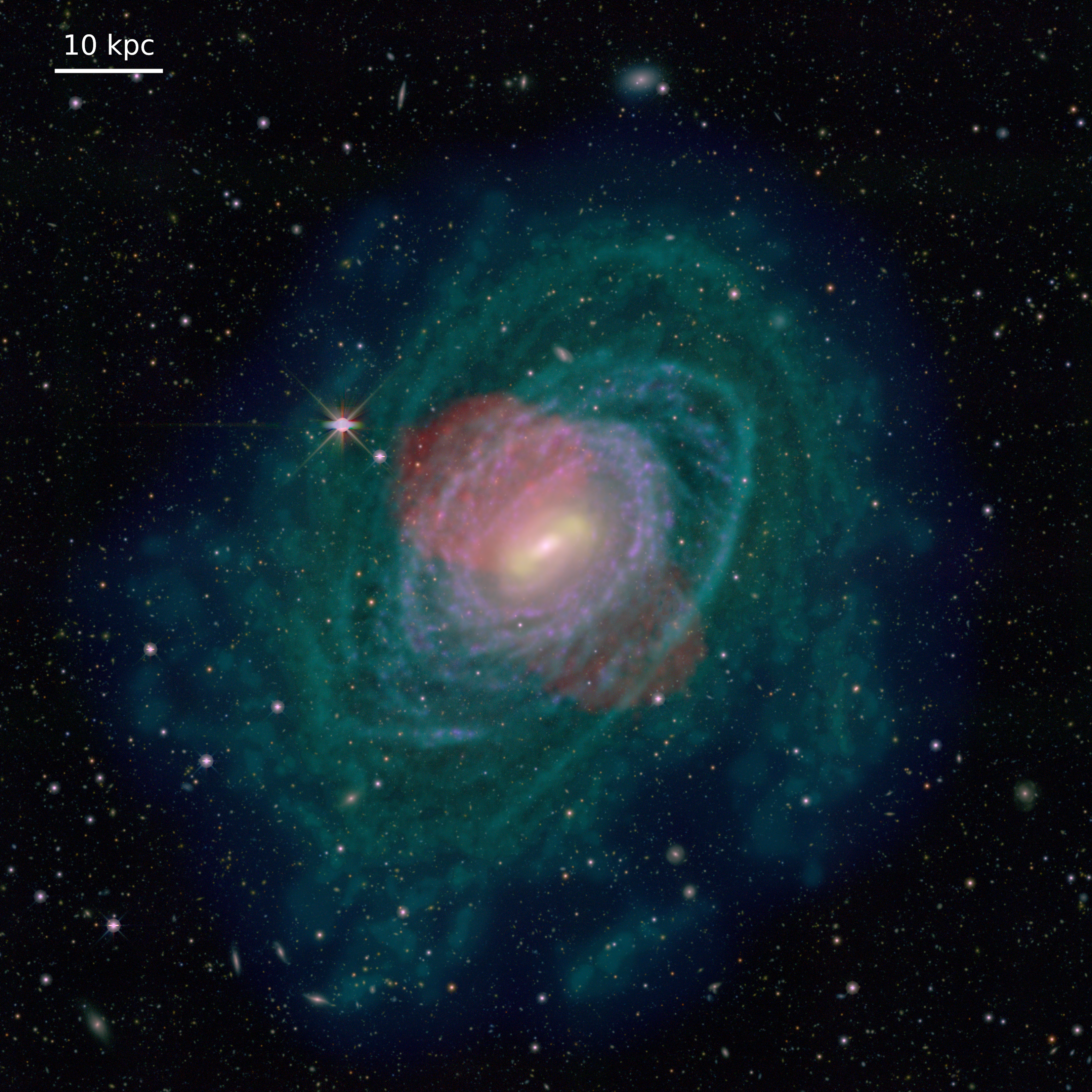}
    \caption{Multiwavelength image of NGC 1371. The background shows the combined $gzri$ optical image from Dark Energy Camera Legacy Survey (DECaLS, \citealt{decals}). The MeerKAT high-resolution 1.4 GHz radio continuum is shown in yellow and red. The UV emission as observed by GALEX is overlaid in pink. The multi-resolution H{\sc i} from the MHOONGOOSE observations is given in green and blue.}
    \label{fig:overlay}
\end{figure*}

\section{Description of the data}\label{sec:data}
\subsection{H{\sc i} observations}\label{sec:hi}
Detailed information on the MHONGOOSE H{\sc i} observations is given in \citet{mhongoose2}. In short, each galaxy is observed for a total of 55 hours, split into ten 5.5-hour sessions. For NGC 1371 we have an additional 5.5 hour track. The eleven tracks underwent the same calibration procedure and were combined as described in \citet{mhongoose2}. For the H{\sc i}, we used the 32k narrow band data, consisting of 32768 channels, each 3.265 kHz wide, giving a total bandwidth of 107 MHz. We binned the channels by 2, leading to a final velocity resolution of 1.4 km s$^{-1}$.
\\\indent The calibration was done with the Containerized Automated Radio Astronomy Calibration (\texttt{CARAcal}) pipeline \citep{caracal}, which provides an all-in-one package to perform standard data reduction steps such as data flagging, calibration, continuum subtraction, and spectral line imaging. \texttt{WSClean} \citep{wscleana,wscleanb} within the \texttt{CARAcal} environment cleaned the H{\sc i} data iteratively using cleaning masks at progressively higher angular resolution, providing six H{\sc i} cubes spanning a range of angular resolution, from $7''$ to $90''$. All cubes have a spectral resolution of 1.4 km s$^{-1}$ and cover a field of view of 1.5$^\circ$.
\\\indent In this work we aim to investigate the effects of stellar and AGN feedback across the disk, thus, we mainly used a high-resolution cube (beam: $13.7''\times9.6''$), created with a robust parameter $r=0.5$ and without u-v tapering. The cube spans a range of $\sim$1000 km s$^{-1}$ centred on 1456.4 km s$^{-1}$, the central velocity of the galaxy and reaches a 3$\sigma$ limiting column density of 1.1 $\times10^{19}$ cm$^{-2}$ over 16 km s$^{-1}$, where $\sigma=1.1\times10^{18}$ cm$^{-2}$ is the noise level.

\subsection{1.4 GHz continuum}\label{sec:cont}
In parallel with the 32k narrow band data, observations were also obtained using the 4k-correlator mode with the same integration times. We use these to create a deep 1.4 GHz continuum image. The 4k data consist of 4096 channels with a channel width of 208.984 kHz, thus providing a bandwidth of 856 MHz centred on a frequency of 1284 MHz. Only the frequency range from 856-1650 MHz was considered (due to the drop-off of sensitivity towards the band edges).
\\\indent The 4k-data were also reduced with \texttt{CARAcal} following standard calibration and flagging procedures. In the 910-950 MHz and 1510-1610 MHz bands all the data were flagged, while in the 1020-1060 MHz, 1080-1320 MHz and 1460-1500 MHz ranges about 60\% of the data per channel were discarded. The 1408-1423 MHz band was additionally flagged to avoid bright H{\sc i} emission. Channels were averaged by a factor five, and this averaged data set was used for three cycles of self-calibration within \texttt{CARAcal}. The resulting continuum measurement sets were combined and deconvolved with the \texttt{oxkat} package \citep{oxcat}. This was also used to apply an additional direction-dependent calibration.
\\\indent The wide bandwidth and long integration times make this observation one of the deeper continuum images made with MeerKAT of nearby galaxies, reaching a final noise level of $2.6\mu$Jy at an angular resolution of $7.4''\times6.7''$ using a robust parameter $r=0$. In Fig. \ref{fig:overlay}, we present a multiwavelength image of NGC 1371, incorporating several of the datasets discussed in this paper.

\subsection{Ancillary data}\label{sec:anc}
Beside the new MeerKAT observations, in this paper we used archival data from other facilities to study star formation and the possible AGN emission. Although H{\sc i} can be seen as the gas reservoir for future star formation, it is molecular hydrogen (H$_2$), formed by cooling of H{\sc i} clouds \citep{clark12,walch15}, which directly fuels stellar production \citep{meidt15,schinnerer19,walter20}. While H$_2$ itself is difficult to observe directly due to its lack of a permanent electric dipole moment, it can be traced by emission of carbon monoxide (CO) \citep{tacconi13,bolatto13,audibert22,audibert23} at millimetre wavelength. Therefore, we used Morita Atacama Compact Array (ACA; \citealt{aca}) observations of NGC 1371 to trace the molecular ISM. These data were taken under the programme ID 2022.1.01314.S (PI: Adam Leroy). The bandwidth covers $\sim500$ km s$^{-1}$ centred on the CO$_{1\to0}$ rest frequency $\nu_{\text{obs}}=115.271$ GHz. The FoV is a $93''\times143''$ (8.9 kpc $\times$13.7 kpc) rectangular mosaic around the major axis and centred on the galaxy.
\\\indent The 1.4 GHz continuum is a tracer of the star formation (e.g., \citealt{cook24}), but it is sensitive to external contamination such as AGN. A more robust estimator of the SFR is therefore obtained by combining UV and infrared emission \citep{leroy08}. As such, we retrieved the Wide-field Infrared Survey Explorer (WISE; \citealt{wise}) 12 $\mu m$ (W3) and the GALEX FUV maps from \citet{leroy19} to quantify the current SFR of NGC 1371.
\\\indent The H{\sc i} and CO emission provides information on the colder phases of the ISM. For the most comprehensive view on what can cause the low SFR in NGC 1371 it is important to study also the hot gas. This is tracked by ionised atomic hydrogen (H$\alpha$) emission. Archival data are, however, restricted to the galactic centre. Indeed, the available H$\alpha$ observations come from integral field spectroscopy data taken by the Multi Unit Spectroscopic Explorer (MUSE; \citealt{muse}). MUSE observed the inner 5 kpc of the galaxy at $\sim$20-pc resolution under programme ID 108.227L.001 (PI: Peter Erwin). We retrieved the reduced data cube from the ESO Science Portal.

\begin{figure*}
    \centering
    \includegraphics[width=\hsize]{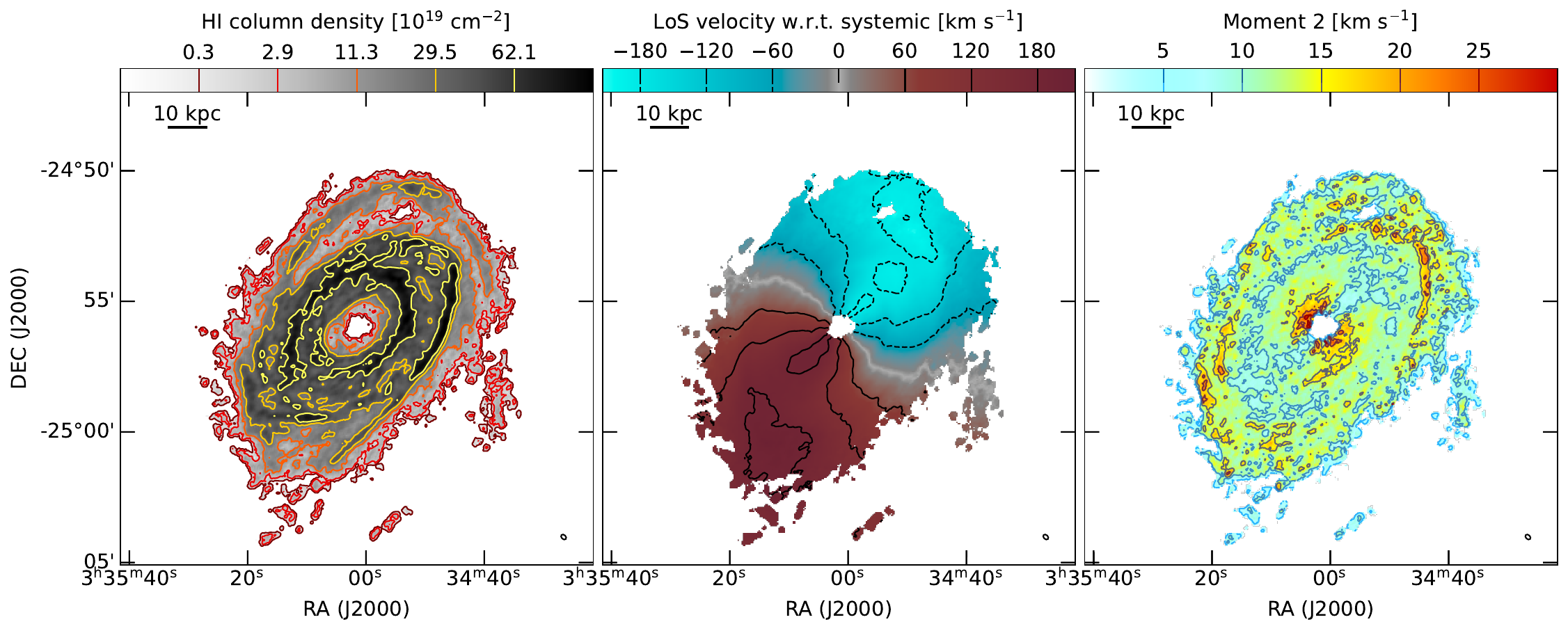}
    \caption{High-resolution H{\sc i} moment maps of NGC 1371. \textit{Left panel}: primary beam-corrected H\,{\sc i} column-density map. The contour levels are given on the colorbar. In the bottom right we show the $13.7''\times9.6''$ beam. The 10-kpc reference scale is shown in the top left. \textit{Central panel}: Intensity-weighted mean velocity field with respect to the systemic velocity (1456.4 km s$^{-1}$). Dashed contours refer to the approaching side, while solid lines correspond to receding velocities. \textit{Right panel}: Second moment map, i.e. the velocity spread along each LoS. It corresponds to the velocity dispersion when the emission comes from a single line, or to the spread of multiple components at different velocities.}
    \label{fig:maps}
\end{figure*}

\begin{figure*}
    \centering
    \includegraphics[width=\hsize]{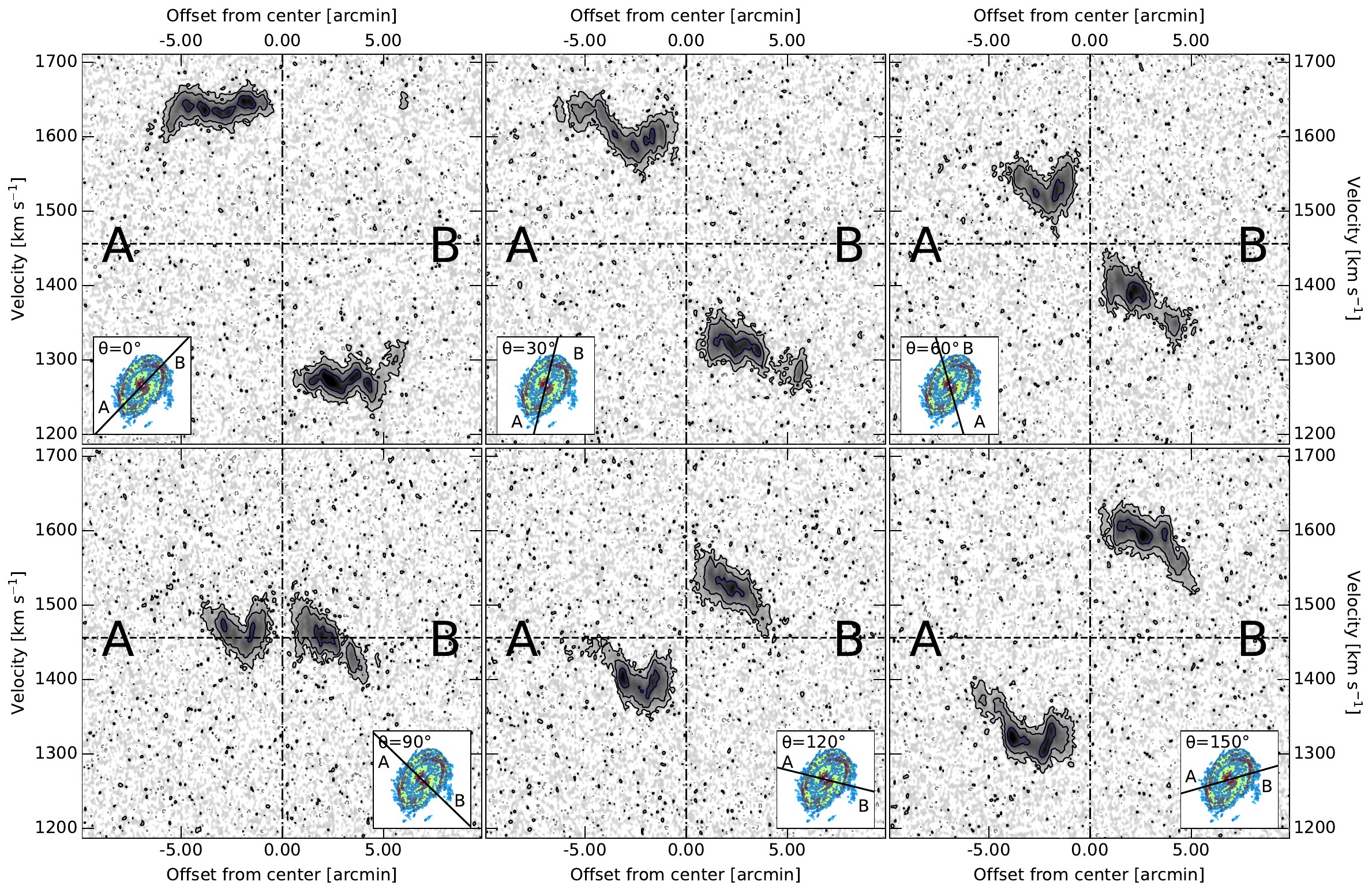}
    \caption{Position-velocity diagrams at azimuthal angles ($\theta$) of (0, 30, 60, 90, 120, 150) degrees with respect to the major axis in the plane of the sky. The data cube was spectrally smoothed with a 7 km s$^{-1}$ Hanning kernel. Solid contour levels denote the (4, 16, 64, 256, 1024)$\sigma$ level, where $\sigma=0.03$ mJy beam$^{-1}$. The dashed grey contours refers to the $-4\sigma$ level. The dashed black horizontal line corresponds to the systemic velocity, while the dashed-dotted black vertical line denotes the centre of the galaxy. The insets shows the moment-2 map and the slice along which the $pv$ is extracted. Letters $A$ and $B$ provides the orientation in the main panels.}
    \label{fig:pv}
\end{figure*}

\begin{figure}
    \centering
    \includegraphics[width=\hsize]{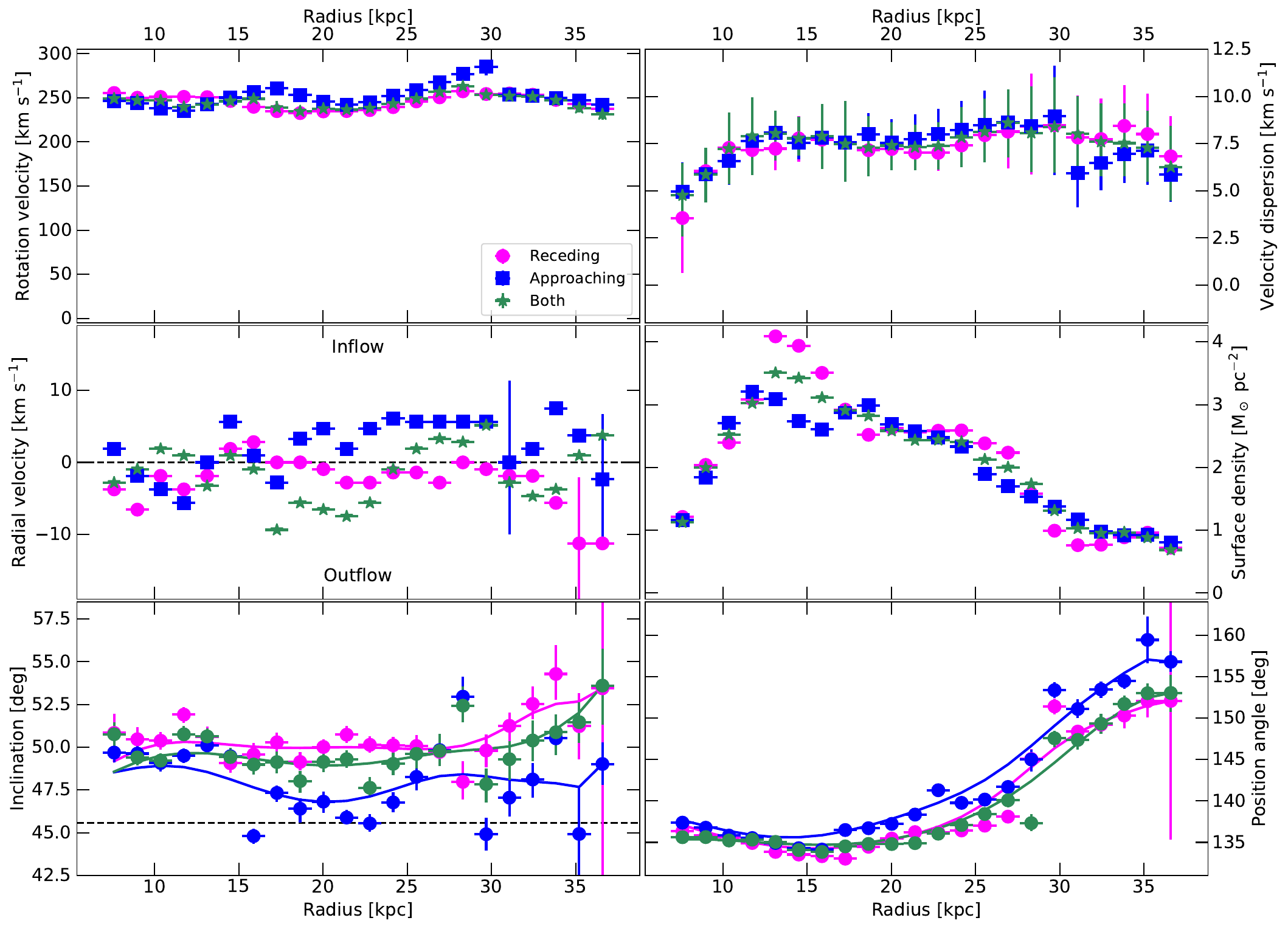}
    \caption{Comparison of the best-fit parameters for the 3D model derived by \texttt{3D-Barolo} fitting the approaching and receding side of the galaxy separately and simultaneously. From left to right and from top to bottom each panel shows $v_\text{rot}$, $\sigma_0$, $v_\text{rad}$, $\Sigma_{\text{H{\sc i}}}$, $i$ and PA. Pink points refer to the best-fit on the receding side, blue to the approaching and green to simultaneous fit. The black horizontal dashed line in the $v_\text{rad}$ panel is a visual guide indicating the 0-level, whereas in the $i$ panel is the indicative optical inclination angle from NED \citep{mhongoose2}.}
    \label{fig:model}
\end{figure}

\begin{figure}
    \centering
    \includegraphics[width=\hsize]{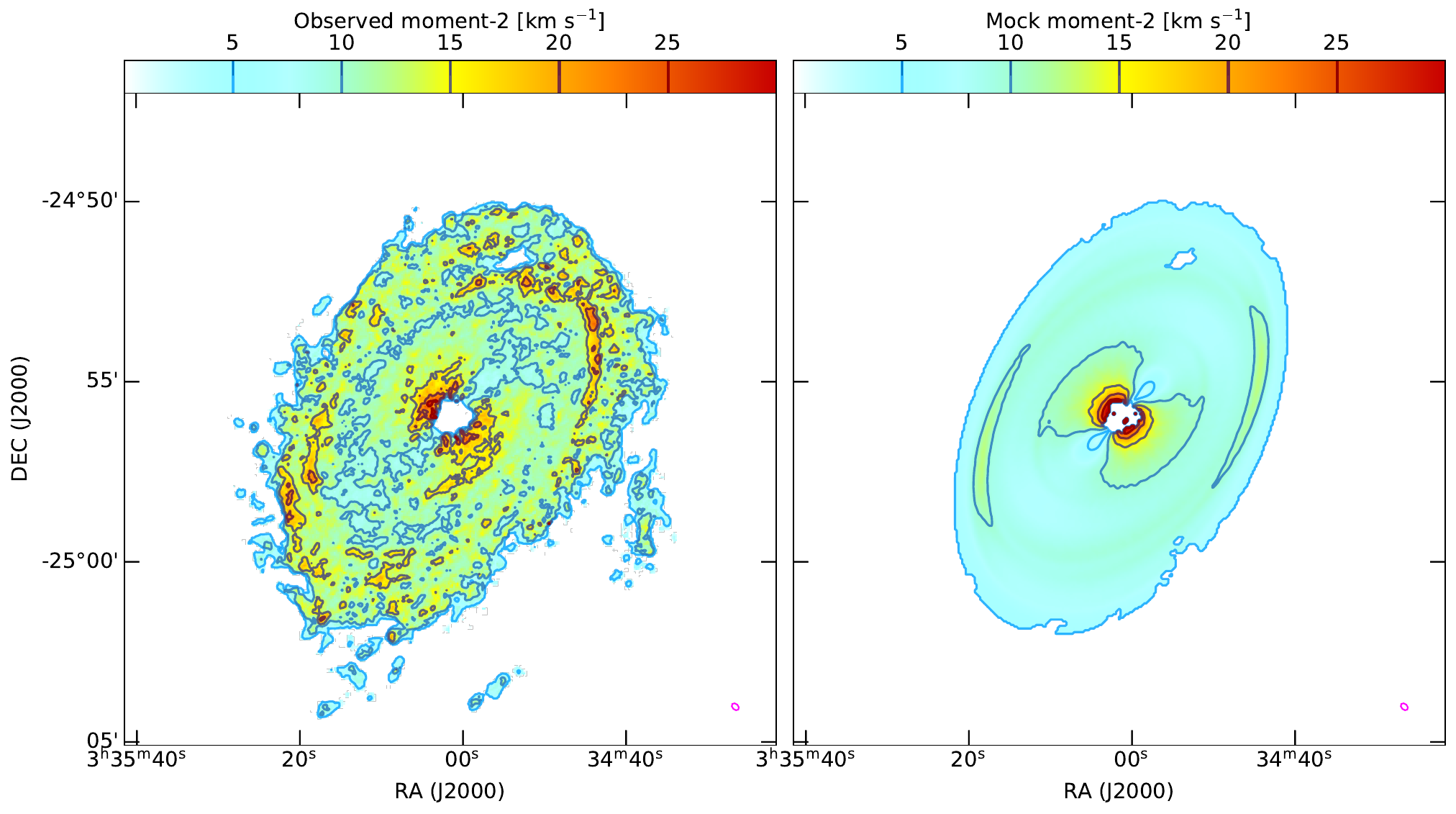}
    \caption{Comparison between the observed (left) and mock (right) moment-2 map derived with \texttt{3D-Barolo}. In the bottom-right corner the $13.7''\times9.6''$ beam is displayed.}
    \label{fig:beam}
\end{figure}

\begin{figure*}
    \centering
    \includegraphics[width=\hsize]{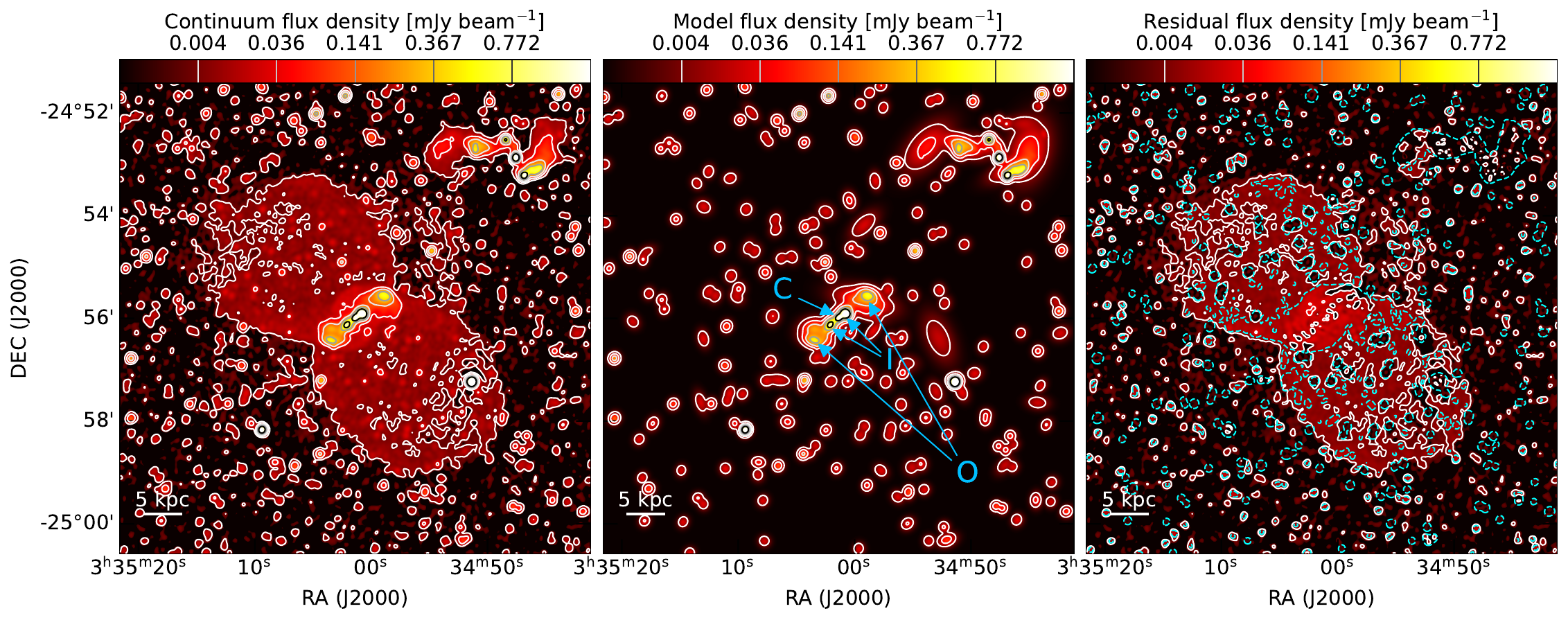}
    \caption{1.4 GHz continuum maps. From left to right: the observed radio continuum, the best-fit model from \texttt{pPyBDSF} and the residuals in terms of data$-$model. The contour levels, with the lowest corresponding to $3\sigma$, are reported on the colorbar at the top. A 5-kpc reference line is given at the bottom-left corner. In the central panel we indicates with blue arrows the location of the core (C), the two inner hotspots (I) and the two outer hotspots (O). In the right-side panel we highlight the edge-brightening of the bubbles with the cyan dashed contours, corresponding to a flux density of $10\mu$Jy beam$^{-1}$.}
    \label{fig:pybdsf}
\end{figure*}

\begin{figure*}
    \centering
    \includegraphics[width=\hsize]{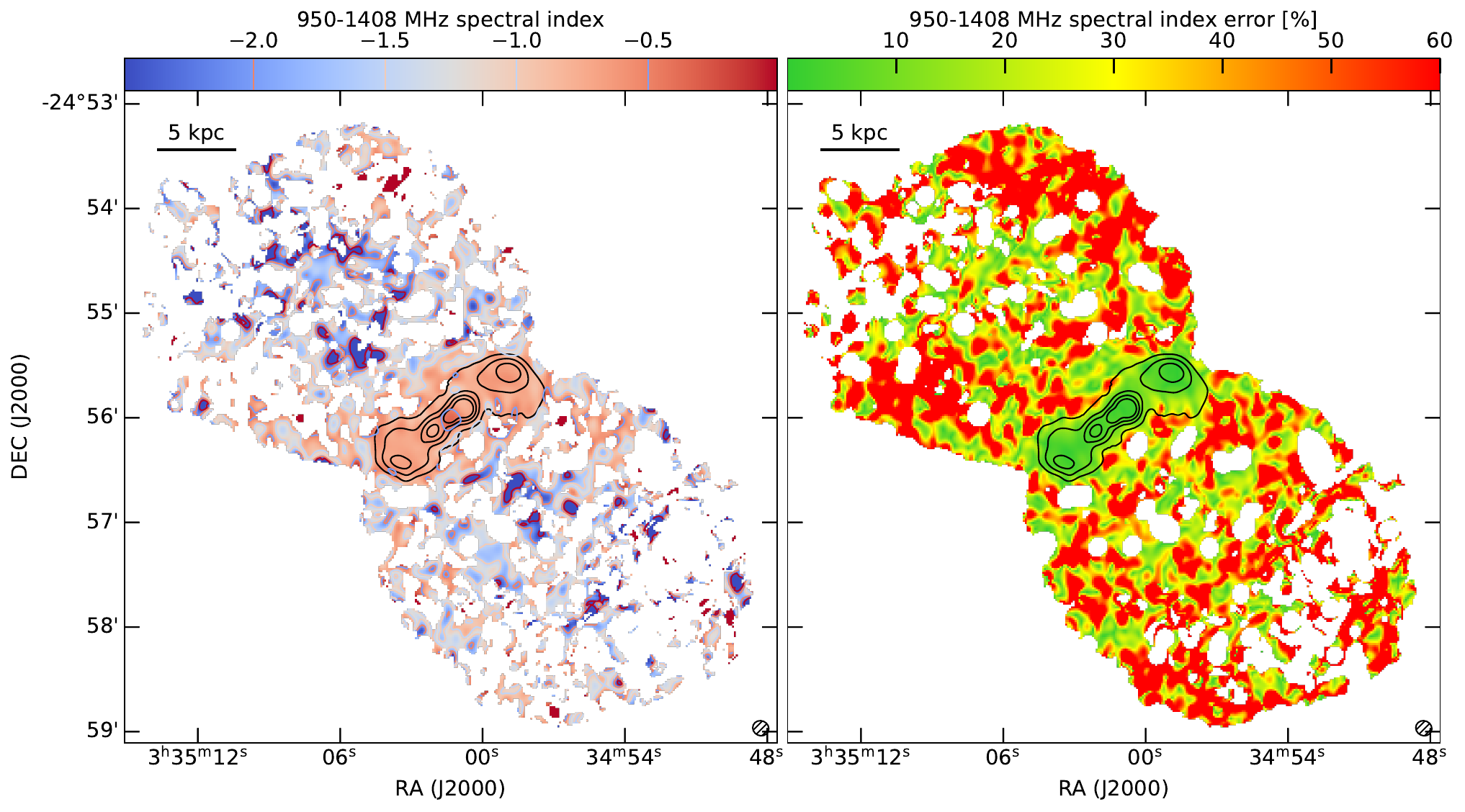}
    \caption{Spatially resolved 950-to-1408 MHz spectral index map (left panel) and percentage uncertainty (right panel). The white regions in the left-side panel are due to the masking of contaminant sources and unreliable fits (see text for the details). The contour levels are given in the colorbar. On the right-side panel we show the percentage error on the spectral index in terms of $100\cdot\frac{\Delta\alpha}{\alpha}$, where $\Delta\alpha$ is the uncertainty on $\alpha$ as derived by the least-square fit. LoS where $\Delta\alpha>50\%$ are marked in red. These were removed from the map on the left. Contaminant sources are masked. In both maps, the black contours refers to the radio continuum from the jet (levels: 36, 141, 367, 772 $\mu$Jy beam$^{-1}$), while the $7.4''\times6.7''$ beam is reported in the bottom-right corner and a 5-kpc reference line is given in the top-left corner.}
    \label{fig:index}
\end{figure*}

\begin{figure}
    \centering
    \includegraphics[width=\hsize]{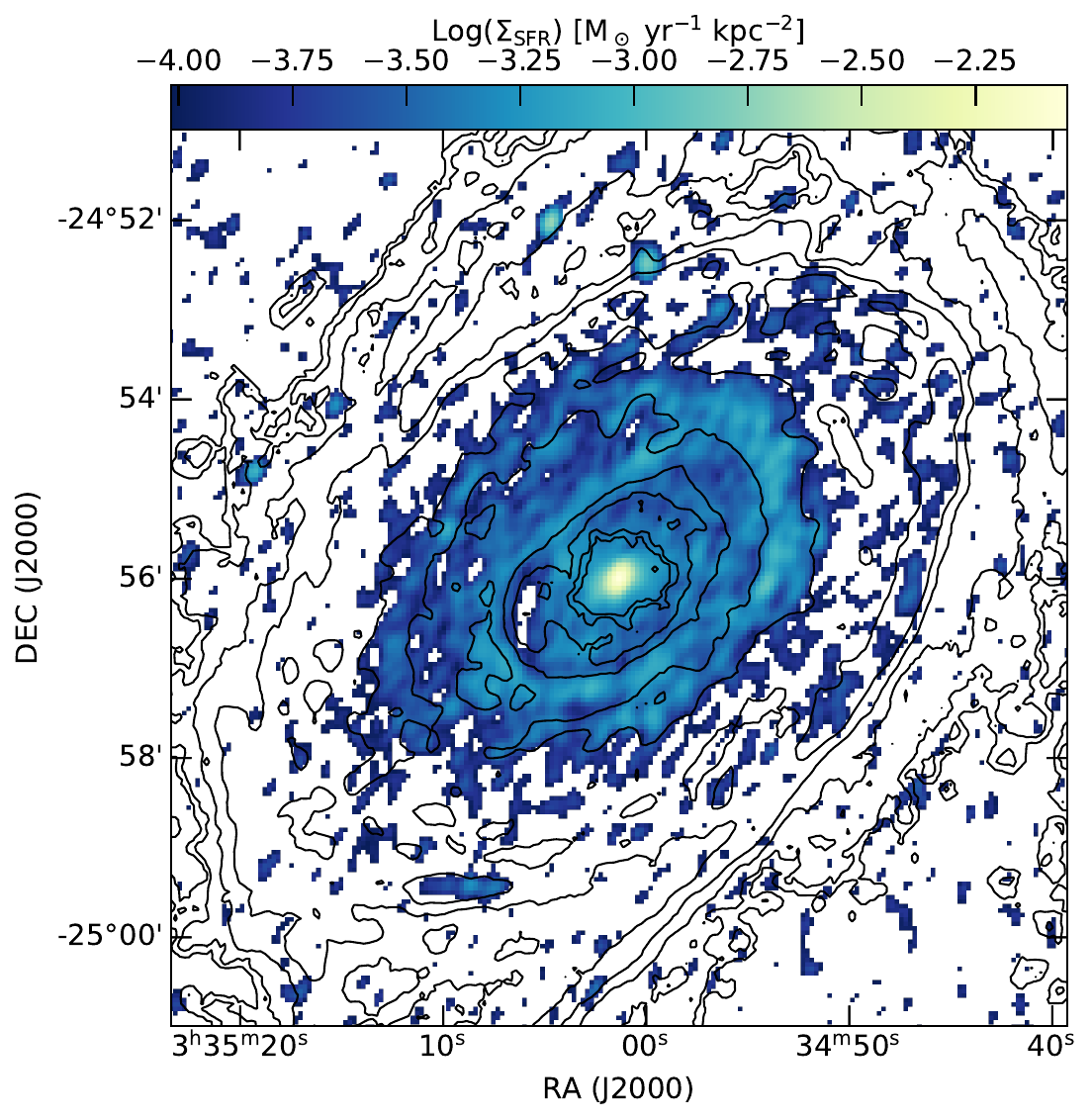}
    \caption{Star formation density map derived with the prescription given in \citet{leroy19}. The black contours correspond to H{\sc i} column densities of (0.3, 2.9, 11.3, 29.5, 62.1) $\times$ 10\textsuperscript{19} cm\textsuperscript{-2}.}
    \label{fig:sfr}
\end{figure}

\begin{figure*}
    \centering
    \includegraphics[width=\hsize]{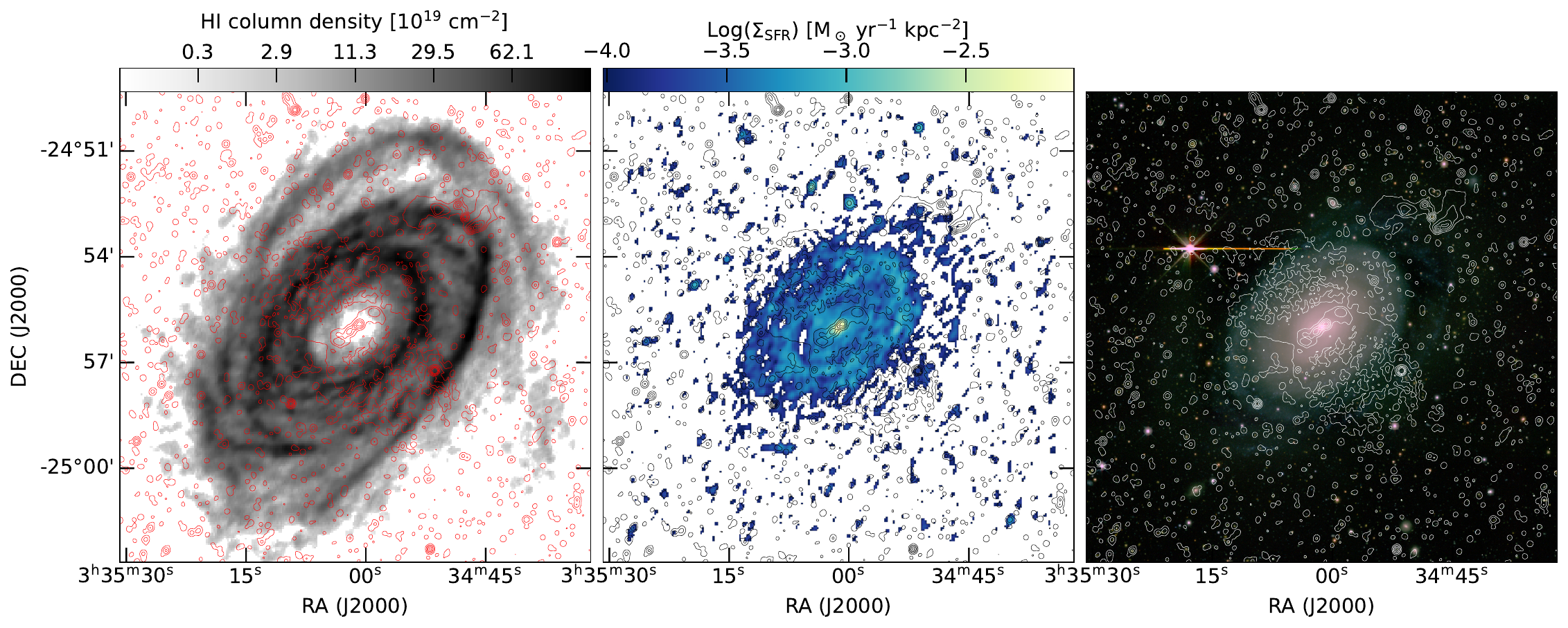}
    \caption{Overlays of the H{\sc i} (left), star formation density (center) and optical (right) map with the radio continuum contours. Contour levels are (0.006, 0.06, 0.2, 0.6, 1.3) mJy beam$^{-1}$, with the lowest being equivalent to 5$\sigma$. The optical image is an RGB composition of the \textit{gzri} DECaLS observations.}
    \label{fig:cont_overlays}
\end{figure*}

\section{Data Analysis}\label{sec:ana}
\subsection{H{\sc i} moment maps}\label{sec:maps}
In Fig. \ref{fig:maps}, we present, from left to right, the $\sim10''$-resolution H{\sc i} intensity (moment-0), intensity-weighted mean velocity\footnote{The velocity field colour map look-up table throughout this paper is taken from \citet{canvas}.} (moment-1), and line-of-sight (LoS) velocity width (moment-2) map for NGC 1371, as derived by the Source Finding Application (\texttt{SoFiA-2}, \citealt{sofia2}) using the parameters described in \citet{mhongoose2}. The H{\sc i} is distributed along tight spiral arms, with the outermost appearing to be disrupted. There is also a central H{\sc i} hole with a diameter of $\sim$5 kpc. This hole is not due to absorption, as there is no significant negative flux within it, but rather to the absence of H{\sc i} at the limiting column density of the map. This remains true even when inspecting the more sensitive, 30$''$ resolution data, which reach a 3$\sigma$ limiting column density of $1.4\times10^{18}$ cm$^{-2}$ over 16 km s$^{-1}$.
\\\indent The moment-1 map shows that most of the H{\sc i} disc is undergoing regular rotation. The kinematic minor axis appears distorted, probably due to the warping of the outer disk. Nevertheless, other processes, such as radial motions and variations in the inclination angle, also affect the minor axis \citep{begeman87,diteodoro21}. The kinematics at the edge of the H{\sc i} hole is regular. In Fig. \ref{fig:chanmap} we show the 1250 km s$^{-1}$ to 1665 km s$^{-1}$ channel maps of the H{\sc i} cube in steps of 28 km s$^{-1}$ to further illustrate the kinematical properties of the disk. These channel maps contain the H{\sc i} intensity at a given velocity, providing an immediate visualization of the gas rotation. The maps also clearly shows the H{\sc i} hole. Furthermore, in Fig. \ref{fig:pv} we include the position-velocity ($pv$) slices extracted\footnote{The $pv$ are 1 beam wide and were extracted with the \texttt{pvextractor} package (\url{https://pvextractor.readthedocs.io/en/latest/}).} along azimuthal angles of (0, 30, 60, 90, 120, 150) degrees, where 0 corresponds to the position angle of the major axis ($\varphi=135^\circ$). The $pv$ diagram is a slice across the cube that gives the velocity of the gas as a function of a spatial coordinate, in this case the distance from the centre of the galaxy. Since the velocity of the gas along the LoS is given by \citep{begeman87}
\begin{equation}\label{eq:vlos}
    v_\text{los}=v_\text{sys}+(v_\text{rot}\cos{\theta}+v_\text{rad}\sin{\theta})\sin{i}
,\end{equation}
where $v_\text{sys}$, $v_\text{rad}$, and $v_\text{rot}$ are the systemic, radial and rotation velocity, respectively, while $\theta$ and $i$ are the azimuthal and inclination angle, the $pv$ along the major axis ($\theta=0^\circ$, top-left panel of Fig. \ref{fig:pv}) traces the rotation curve of the galaxy. Instead, along the minor axis ($\theta=90^\circ$, bottom-left panel of Fig. \ref{fig:pv}), non-circular motions can be identified via deviations from the systemic velocity. Intermediate values of $\theta$ allow for a complete overview of the gas kinematics.
\\\indent In the upper-right corner of the top-left panel of Fig. \ref{fig:pv} a significant clumpy emission is visible at $\sim6'$ from the galactic centre at a velocity of $\sim 1650$ km S$^{-1}$. The H{\sc i} mass of this object, that we will refer to as MKT 033442-245148 ($\alpha=3^h34^m42.31^s$, $\delta=-24^d51^m47.52^s$, $cz=1647$ km s$^{-1}$), is $M_{\rm H{\sc i}}=6.6\times10^6$ M$_\odot$. The overlay between the H{\sc i} contours and the DECaLs $gzri$ image presented in Fig. \ref{fig:clump} reveals that MKT 033442-245148 is likely a dwarf galaxy. Searching for known objects at the location of MKT 033442-245148 in the SIMBAD Astronomical Database \citep{simbad} and in the NASA/IPAC Extragalactic Database\footnote{The NASA/IPAC Extragalactic Database (NED) is funded by the National Aeronautics and Space Administration and operated by the California Institute of Technology.} leads to no results. Therefore, MKT 033442-245148 is likely a new galaxy. An immediate question raising from this finding is whether this dwarf is interacting with NGC 1371. The moment maps of MKT 033442-245148, provided in Fig. \ref{fig:clumpmaps}, show no clear signatures of tidal tails or gas stripping. The morphology and kinematics are fairly regular, with the only prominent feature being an asymmetry in the NE side of the galaxy. However, this asymmetry is also marginally resolved, therefore we do not consider it a strong indication of ongoing interaction between NGC 1371 and MKT 033442-245148. The above considerations holds even when inspecting the 30$''$ resolution data, suggesting that the $>10^{18}$ cm$^{-2}$ H{\sc i} in this galaxy is relaxed.
\\\indent The moment-2 map of NGC 1371 shows $\sim15$ km s$^{-1}$ values throughout the disk and $>15$ km s$^{-1}$ around the H{\sc i} hole along the minor axis of the disc and between the two spiral arms. However, interpreting these values is not straightforward, as they represent the velocity spread along the LoS: a high value can result from either a single structure with a large velocity width or multiple (narrow) components distributed along the LoS. To investigate this, we performed a detailed kinematical modelling of the H{\sc i} disc with the 3D-Based Analysis of Rotating Objects via Line Observations (\texttt{3D-Barolo}, \citealt{barolo}).
\\\indent The modelling is performed via `tilted-ring' fitting \citep{begeman87}, where the disc is approximated in concentric rings. Each ring is parametrised with a set of five geometrical variables, namely, the rotation centroid ($x_0$, $y_0$), the position angle on the sky (PA), the inclination angle ($i$) and the scale height ($z_0$), and four kinematical parameters, that are the systemic, rotation, and radial velocity ($v_\text{sys}$, $v_\text{rot}$, $v_\text{rad}$) and the intrinsic velocity dispersion ($\sigma_0$) as a function of radius. For this fit we used the \texttt{3DFIT} task of \texttt{3D-Barolo} on the H{\sc i} disc divided into rings with the width along the major axis equal to the beam (13.7$''$). The fitting procedure is a two-step process, described in detail in \citet{diteodoro21}, where in the first step we constrained the rotation curve and in the second we fit the radial velocities (if present). This procedure is done on the approaching and receding side separately, as well as both sides simultaneously. To constrain the rotation curve we fixed the scale-height of the disc to 100 pc. Using different values is not critical: as long as they are between 0 and 1 kpc, differences in the best-fit are negligible. As from Eq. (\ref{eq:vlos}) the radial velocities are minimal close to the major axis ($\theta=0^\circ$), whereas $v_\text{rot}$ has the larger contribution to $v_\text{los}$, we also fixed $v_\text{rad}=0$ km s$^{-1}$ and set the weighting function for the fit to $\cos^2(\theta)$, such that the fit will focus on minimizing the residuals along the major axis \citep{barolo}. The minimization is done on $|\text{model}-\text{data}|$ to balance the residuals of the bright and faint emission \citep{barolo}. The free parameter of the fit are, therefore, $x_0$, $y_0$, $v_\text{rot}$, $v_\text{sys}$, PA, $i$ and the intrinsic dispersion $\sigma_0$. This first step consists furthermore in a two-stage fit. In the first stage, the free parameter are fitted. However, a systematic issue in the 3D modelling is that the fits of PA and $i$ often show unrealistic scatter \citep{barolo}. Thus, in the second stage of the fit \texttt{3D-Barolo} regularises PA and $i$ and repeats the fit on the free parameters. We chose Bezier functions \citep{barolo} for this regularisation, as we noticed that the radial profile of PA and $i$ cannot be approximated by a constant or a simple polynomial. The result of this first, two-stage fit is used as prior for the second step, where we want to fit for radial motions. Equation (\ref{eq:vlos}) tells that the effect of $v_\text{rad}$ is maximised along the minor axis ($\theta=90^\circ$), whereas $v_\text{rot}$ is now minimal. Consequently, any small error in the determination of $v_\text{rot}$ has a minor impact in this step. We kept the best-fit values from the first step fixed and let $v_\text{rad}$ free to vary. We also changed the weighting function to $\sin^2(\theta)$, as we want to minimize the residuals along the minor axis. Furthermore, in this step only the first stage of the fit is used, as PA and $i$ are fixed.
\\\indent In Fig. \ref{fig:model} we show the comparison of the radial profiles of the geometrical and kinematical parameters as derived by fitting only the receding (in magenta) and approaching side (in blue) or both sides simultaneously (in green). The differences in the best-fits between the two sides are negligible. The jump in $v_\text{rot}$ and $\sigma_0$ at 30 kpc correspond to the location of the outer spiral arm. Similarly, the difference in the H{\sc i} surface density around 15 kpc is due to the asymmetry in the H{\sc i} distribution of the inner disk. The parameter having the largest scatter among the three models is $i$. However, an unambiguous estimate for $i$ is generally hard to achieve even with high-resolution, high S/N data (e.g., \citealt{diteodoro21,pina22,pina24}). Furthermore, comparing the model with the data using the same $pv$ slices of Fig. \ref{fig:pv} (see Fig. \ref{fig:pvboth}, \ref{fig:pvrec} and \ref{fig:pvapp} in Appendix \ref{app:barolo}) we found always a good agreement with the data.
\\\indent Hereafter we will consider as the best-fit model the one derived from the simultaneous fit of both sides. This gives a constant centroid at $(x_0,y_0)=(3^{h}\:35^{m}\:01.56^{s},-24^d\:56^m\:00.12^s)$, almost identical to the fiducial position of $(x_0,y_0)=(3^{h}\:35^{m}\:01.35^{s},-24^d\:55^m\:59.58^s)$ listed in NED from WISE observations, and a constant systemic velocity of $v_{sys}=1456.4$ km s$^{-1}$, the same value derived from the global profile \citep{mhongoose2}. This already indicates that the kinematics of this galaxy is gloablly regular, as there are no major disturbances able to shift the central velocity of the global profile away from $v_\text{sys}$. Indeed, the radial velocities are consistent with 0 km s$^{-1}$ over the whole disc ($\langle v_\text{rad}\rangle=-1.5\pm3.9$ km s$^{-1}$). In Fig. \ref{fig:beam}, we compare the observed moment-2 map with the one derived by \texttt{3D-Barolo} using the aforementioned model. The high dispersion in the galactic centre is also present in the mock moment map, indicating that it is a resolution effect caused by the extremely steep rise on the rotation curve compared to the beam size.

\subsection{Radio continuum}\label{sec:radio}
In the left-hand panel of Fig. \ref{fig:pybdsf} we present the radio continuum map. We detected\footnote{The north-west emission is a background AGN.} a previously known jet-like structure \citep{omar05,grundy23}, comprising of a core, two inner hotspots, and two outer hotspots. Due to the high sensitivity of MeerKAT, we also detect two (in projection) $\sim10$-kpc-wide nuclear bubbles that have never been observed before in this galaxy. We used the \texttt{process\_image} task of the Python Blob Detector and Source Finder\footnote{We ran the task with its default input arguments, except for \texttt{thresh\_isl}, which was set to 2.5.} (\texttt{PyBDSF}; \citealt{pybdsf}) to quantify the 1.4 GHz luminosity ($L_{\text{1.4 GHz}}$) of these structures and the galaxy as a whole. \texttt{PyBDSF} decomposes an image into a set of Gaussians and subsequently merges them into sources. These sources are then parametrised and recorded in a catalogue. The software also produces model and residual images for visual inspection, shown in the central and right-hand panels of Fig. \ref{fig:pybdsf}, respectively. The total luminosity of the entire jet-like structure, as derived by \texttt{PyBDSF}, is $L^{\text{jet}}_{\text{1.4 GHz}}=(9.75\pm0.04)\times10^{20}$ W Hz$^{-1}$, consistent with the measurements from the literature summarised in Table 1 of \citet{grundy23}. The core alone has a luminosity of $L^{\text{core}}_{\text{1.4 GHz}}=(7.17\pm0.05)\times10^{19}$ W Hz$^{-1}$.
\\\indent From the residual image, we also estimated the luminosity of the bubbles. We isolated them by applying an initial $3\sigma$ cut to the residual image (right panel of Fig. \ref{fig:pybdsf}), followed by manual masking of the remaining low-level sources outside the bubbles. Summing the flux in the map yields $L^{\text{bubbles}}_{\text{1.4 GHz}}<4.6\times10^{20}$ W Hz$^{-1}$, where the upper limit is to account for contamination by low-level background sources and residual faint emission from the jets or star forming regions not included in the \texttt{PyBDSF} modelling. The total continuum luminosity of NGC 1371 is therefore $L_{\text{1.4 GHz}}<1.5\times10^{21}$ W Hz$^{-1}$, placing the galaxy at the faint end of the 1.4 GHz luminosity function of radio galaxies, both observationally \citep{pracy16,buttler19} and theoretically \citep{thomas21}. In Table \ref{table:power} we list the 1.4 GHz flux and luminosity for the various radio components.
\\\indent The high SNR of our data allows us to derive the spatially resolved distribution of the spectral index ($\alpha$) in the 950–1408 MHz band, defined by the non-thermal power law $S_\nu\propto\nu^\alpha$, where $S_\nu$ is the monochromatic flux at frequency $\nu$. We convolved the cube to the lowest-resolution beam and to avoid contamination of background sources, we used the \texttt{PyBDSF} model image (see the central panel of Fig. \ref{fig:pybdsf}) to mask the LoS containing background quasars. Finally, we determined $\alpha$ using a fit\footnote{The fit is performed with the \texttt{curve\_fit} task of \texttt{SciPy} \citep{scipy}, which performs a non-linear least squares fit to the data.} to the logarithmic spectrum at each position. For each unmasked LoS, only channels with flux $>3\sigma$ were considered for the fit. The uncertainty on $\alpha$ ($\Delta\alpha$) is computed from the covariance matrix of the best-fit parameters. Its distribution in terms of $\frac{\Delta\alpha}{\alpha}$ is given in the right-side panel of Fig. \ref{fig:index}. LoS for which $\frac{\Delta\alpha}{\alpha}>0.5$ are shown in red and were further rejected from the spectral index map, presented in the left-side panel of Fig. \ref{fig:index}.

\subsection{Star formation rate density}\label{sec:sfr}
The high spatial resolution of the MHONGOOSE observations allows for spatially resolved study of the correlation between the H{\sc i} and radio continuum distribution with the SFR. Therefore, we computed SFR density maps ($\Sigma_{\text{SFR}}$) from the WISE W3 and the GALEX FUV data from \citet{leroy19}, which already include masks for foreground stars. After correcting for the inclination of NGC 1371, we converted the FUV ($I_{\text{FUV}}$) and W3 ($I_{\text{W3}}$) intensities into $\Sigma_{\text{SFR}}$ using the following equation, described in \citet{leroy19}:
\begin{equation}\label{eq:sfr}
    \begin{aligned}
    \frac{\Sigma_{\text{SFR}}}{\text{M$\odot$ yr$^{-1}$ kpc$^{-2}$}}=&1.04\times10^{-1}\left(\frac{C{\text{FUV}}}{10^{43.35}}\right)\left(\frac{I_{\text{FUV}}}{\text{1 MJy sr$^{-1}$}}\right)+\\&
    3.77\times10^{-3}\left(\frac{C_{\text{W3}}}{10^{42.79}}\right)\left(\frac{I_{\text{W3}}}{\text{1 MJy sr$^{-1}$}}\right)
    ,\end{aligned}
\end{equation}
where $C_{\text{FUV}}=10^{43.42}$ and $C_{\text{W3}}=10^{42.79}$ are the FUV and W3 correction factors, respectively. The resulting map is shown in Fig. \ref{fig:sfr}.
\\\indent In Fig. \ref{fig:cont_overlays} we overlay the radio continuum with the H{\sc i} moment 0, the $\Sigma_{\text{SFR}}$ map and the optical image from DECaLS. This immediate positional comparison shows that there is little-to-no spatial correlation between the radio continuum and the H{\sc i} distribution. The lack of significant radio emission from the spiral arms is also qualitatively consistent with the low SFR of this galaxy and with the sparse distribution of star-forming regions in the UV imaging. Interestingly, the jet-like structure has no clear counterpart in both the $\Sigma_{\text{SFR}}$ and optical image.

\section{Discussion}\label{sec:disc}
By analysing the H{\sc i} and radio continuum emission, as well as $\Sigma_{\text{SFR}}$ of NGC 1371, we found a jet-like structure within the central H{\sc i} hole having a luminosity of $L^{\text{jet}}_{\text{1.4 GHz}}=(9.75\pm0.04)\times10^{20}$ W Hz$^{-1}$. This structure is not visible in the $\Sigma_{\text{SFR}}$ and is more extended than the stellar bar. In the radio continuum we also detected previously unknown bipolar nuclear bubbles with a projected size of $\sim10$ kpc and luminosity of $L^{\text{bubbles}}_{\text{1.4 GHz}}<4.6\times10^{20}$ W Hz$^{-1}$. In the following we discuss whether those features are attributed to the stellar bar or to an AGN or to both, and how they could explain the low SFR of the galaxy.

\subsection{Is NGC 1371 an active galaxy?}\label{sec:agn}
Whether NGC 1371 hosts an AGN is a matter of debate (see the discussion in Sect. \ref{sec:intro}). Our radio continuum data reveals the presence of a jet-like structure, already detected previously \citep{omar05,grundy23}, that we can now study in more detail to understand if it is coming from an AGN or from the star formation in the stellar bar.
\\\indent One way to distinguish between the two origins is to compare $L_{\text{1.4 GHz}}$ with the SFR, as there is a well-known correlation between $L_{\text{1.4 GHz}}$ and the SFR \citep{cook24}. From the $\Sigma_{\text{SFR}}$ map derived earlier we measured a total SFR within the jet-like region of $\sim0.03$ M$_\odot$ yr$^{-1}$. The SFR-$L_{\text{1.4 GHz}}$ relation from \citet{cook24} gives an estimation for $L_{\text{1.4 GHz}}$ of $1.6\times10^{19}$ W Hz$^{-1}$. This value is approximately 2 dex lower than $L^\text{jet}_{\text{1.4 GHz}}$ derived by \texttt{PyBDSF}. This suggests that star formation is unlikely to be responsible for the jet-like structure and that a low-luminosity AGN is the most probable cause.
\\\indent Another parameter commonly used for distinguishing radio emission from star formation from that of AGN activity is the spectral index. The spectral index associated with star formation typically lies within the range $-0.6$ to $-0.8$ \citep{magnelli15,tabatabaei17,klein18,an21,an24}, whereas AGNs can exhibit a broader range of values \citep{eckart86,zaja19}: spectral indices of $\alpha<-0.7$ are characteristic of optically thin synchrotron emission from cool electrons in radio lobes, while $-0.7<\alpha<-0.4$ indicates a mixed contribution of optically thin and self-absorbed synchrotron emission from jets and values of $\alpha>-0.4$ are typical of AGN cores, where synchrotron self-absorption becomes significant. From the spectral index map presented in Fig. \ref{fig:index} we measured an average spectral index in the radio core of $\alpha=-0.43\pm0.01$, consistent with the synchrotron self-absorption emission from an AGN core, while the radio emission from the jet has $\alpha<-0.7$, which is consistent with the expected value for radio lobes. However, we must point out that our spectral index analysis has some caveats: the derivation of the spectral index strongly depends on the frequency range, the thermal contamination and the acceleration mechanism for the charged particles. Nonetheless, with the further support of the radio excess in the jet-like structure and the lack of similar features in the $\Sigma_{\text{SFR}}$ map and the optical image, as shown in Fig. \ref{fig:cont_overlays}, we conclude that NGC 1371 is an active galaxy and the jet-like structure is tracing the AGN jet, rather than the star formation in the stellar bar. 

\subsection{How common are the bubbles?}\label{sec:bubbles}
Observations of disc galaxies hosting a radio AGN are rare \citep{ledlow98,abdo09,morganti11,doi12,bagchi14,singh15}. Kiloparsec-scale radio bubbles misaligned with the jets are even rarer, with only few known cases up to date (see Table 4 of \citealt{hota06} and also Mrk 231, \citealt{ulvestad99}; M87, \citealt{owen00,degasperin12}; The Teacup, \citealt{harrison15}; NGC 4438, \citealt{li22} and M106, \citealt{zeng23}). Apart from The Teacup (undetermined morphology) and M87 (cD0-1 pec), all the aforementioned galaxies are classified as disc. Regarding the radio luminosity, NGC 4051 and M51 have $L_{\text{1.4 GHz}}$ one dex lower than NGC 1371 \citep{hota06}, but their bubbles are also much smaller, with an estimated extent of $~0.5$ kpc. NGC 3367 has $\sim15.3$ kpc bipolar bubbles\citep{hota06} but its radio emission ($L_{\text{1.4 GHz}}=1.12\times10^{22}$ W Hz$^{-1}$) is much brighter than NGC 1371. The other galaxies have $L_{\text{1.4 GHz}}$ two to three orders of magnitude higher than NGC 1371 and bubble sizes in the range of 1-to-3.4 kpc. Therefore, the bubbles of NGC 1371 have unique aspects.

\begin{table}
    \caption{NGC 1371 radio flux.}
    \label{table:power}
    \centering
    \begin{tabular}{c c c}
    \hline\hline
    Component & Flux & Luminosity \\
    & [mJy] & [W Hz$^{-1}$]\\
    \hline
    Jet               & $15.81\pm0.07$ & $(9.75\pm0.04)\times10^{20}$ \\
    Core              & $1.16\pm0.01$  & $(7.17\pm0.05)\times10^{19}$ \\
    Bubbles           & $<7.4$           & $<4.6\times10^{20}$ \\
    \hline
    \end{tabular}
\end{table}

\begin{table}
    \caption{Comparison of Sgr A$^\star$ and the SMBH in NGC 1371.}
    \label{table:smbh}
    \centering
    \begin{tabular}{c c c c}
    \hline\hline
    & Sgr A$^\star$ & NGC 1371\\
    \hline
    M$_{\text{SMBH}}$ [M$_\odot$]   & $4.14\times10^{6}$  & $10^{7.4}$\\
    L$_\text{bol}$ [erg s$^{-1}$]   & $10^{36}$           & $8\times10^{41}$ \\
    $\dot{m}$ [M$_\odot$ yr$^{-1}$] & $10^{-8}$-$10^{-9}$ & $1.4\times10^{-4}$ \\
    \hline
    \end{tabular}
    \tablefoot{M$_{\text{SMBH}}$ and L$_\text{bol}$ for Sgr A$^\star$ were taken from \citet{eht22a} and \citet{eht22b} and for NGC 1371 from \citet{cisternas13}. The values for $\dot{m}$, instead, were retrieved from \citet{sharma07,eht22a} and \citet{ressler23} for Sgr A$^\star$ and computed by us for NGC 1371.}
\end{table}

\begin{figure}
    \centering
    \includegraphics[width=\hsize]{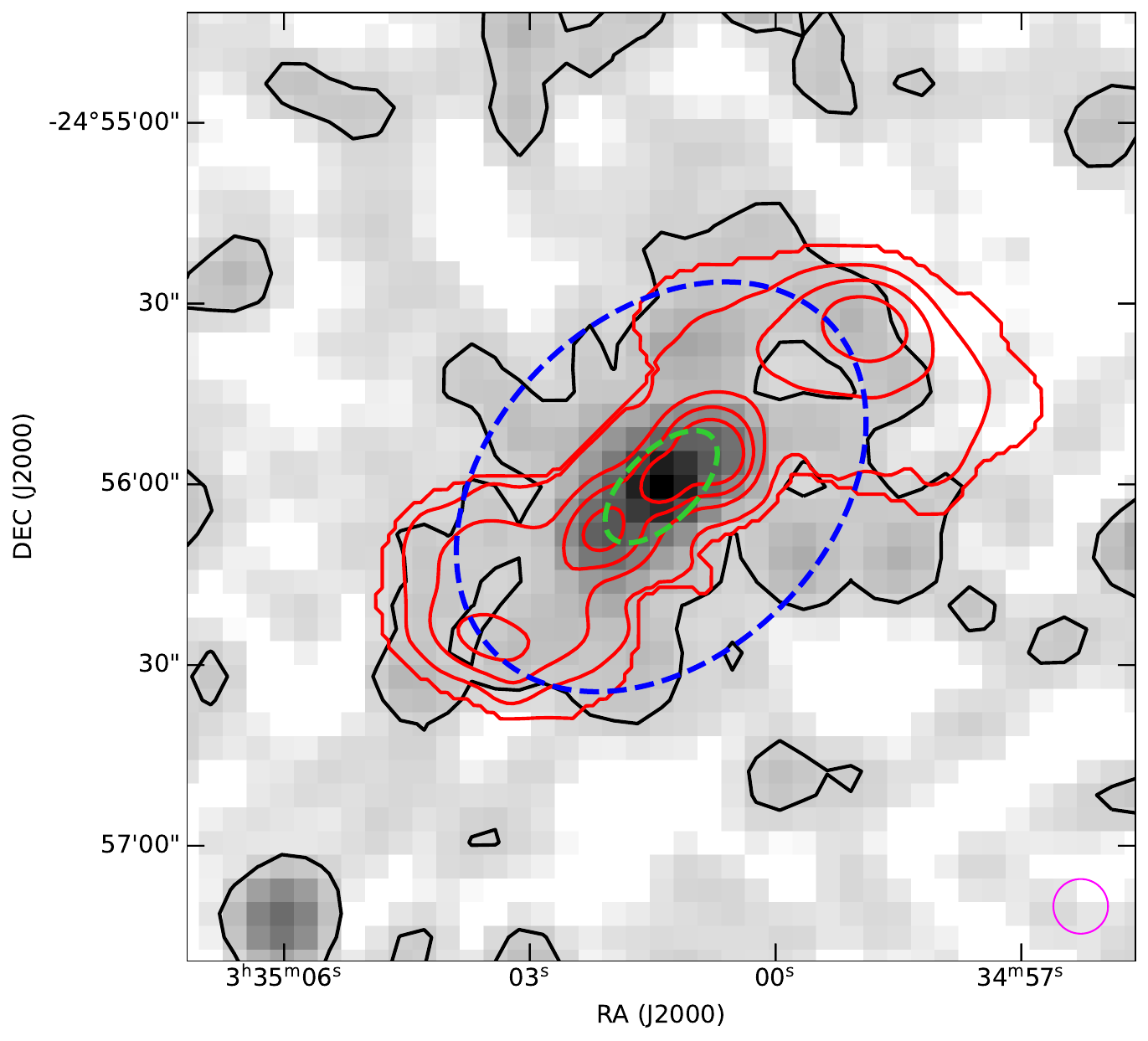}
    \caption{Chandra X-ray image of NGC 1371. The X-ray emission has been smoothed with a Gaussian 1 kpc $\times$ 1 kpc kernel (given in the bottom-right corner) to enhance the faint emission. Black contour denotes the 3$\sigma$ level of the smoothed X-ray image. Red contours refers to the radio continuum (levels: 0.08, 0.31, 0.80, 1.68 mJy beam$^{-1}$, where the beam is $7.4''\times6.7''$). The dashed green and blue ellipses correspond to the extraction regions of the X-ray spectra studied by \citet{hughes07}.}
    \label{fig:xray}
\end{figure}

\subsection{Comparison with the Fermi/eROSITA bubbles}\label{sec:fermi}
In our Galaxy 10-kpc bipolar nuclear bubbles have been detected at $\gamma$- and X-ray energies (reviewed by \citealt{sarkar24}). The radio emission from the so-called `Fermi/eROSITA bubbles' (FEBs) exhibits a northern loop whose outer edge closely follows the FEBs \citep{haslam82,kalberla05,planck16}. There is an ongoing debate on whether this radio emission results from a nearby ($\sim200$ pc away from us) supernova (e.g., \citealt{vidal15,dickinson18}) or nuclear activity (e.g., \citealt{mou14,mou15,zhang20,yang22,mondal22}). Arguments supporting the supernova origin arise from the absence of a southern FEBs radio counterpart, though some simulations suggest this could also result from nuclear activity interacting with an inhomogeneous CGM (e.g., \citealt{sarkar19}). Can the FEBs provide insights into the properties of the bubbles in NGC 1371 and vice versa? In other words, could the radio bubbles in NGC 1371 be extragalactic analogues of the FEBs?
\\\indent Unfortunately, there is no detected X-ray emission for the NGC 1371 bubbles (see Fig. \ref{fig:xray}), preventing a direct comparison. However, similarities can still be explored. \citet{hughes07} studied the X-ray distribution from NGC 1371 and found a compact source coincident with the nucleus and well described by an absorbed power-law spectrum with unabsorbed 0.5-10 keV flux of $F_\text{2-10 keV}=1.2\times10^{-12}$ erg cm$^{-2}$ s$^{-1}$. They also found diffuse emission in a 2.6 kpc $\times$ 1.3 kpc region around the galactic centre, coinciding with the radio jet (see the green ellipse in Fig. \ref{fig:xray}). The X-ray spectrum is consistent with thermal emission from a hot ($kT\sim10$ keV) and a warm ($kT\sim0.3$ keV) medium. The authors suggested that this emission originates from sources linked to recent star formation, but we argue that the X-rays likely trace the ISM heated by the jet, as they correlate well with the brightest radio continuum. \citet{hughes07} detected also very faint and soft diffuse emission ($E<2$ keV) within an 8.6 kpc $\times$ 6.2 kpc region around the AGN (see the blue ellipse in Fig. \ref{fig:xray}), enclosing the radio lobes and the base of the bubbles. Notably, they report a medium temperature of $\sim0.4$ keV, similar to the FEBs ($\sim0.3$ keV; \citealt{kataoka13,miller16,ursino16,kataoka18}).
\\\indent Since NGC 1371 is an active galaxy, it is possible that the bubbles are produced by its AGN. For this we further compare the nuclear activity of NGC 1371 with our Galaxy under the assumption that the FEBs are also ejected by our central SMBH, Sgr A$^\star$. In fact, whether FEBs result from nuclear starburst or AGN activity remains debated (see \citealt{sarkar24} and references therein). Sgr A$^\star$ has a mass of $4.14\times10^6$ M$_\odot$ \citep{eht22a} (see also \citealt{gillessen09,davis14}), current accretion rate of $10^{-8}$-$10^{-9}$ M$_\odot$ yr$^{-1}$ \citep{sharma07,eht22b,ressler23}, and current bolometric luminosity of $\sim10^{36}$ erg s$^{-1}$ \citep{eht22b}. In Table \ref{table:smbh} we compare these quantities for Sgr A$^\star$ and the SMBH in NGC 1371. Estimates in the literature for the mass of the SMBH in NGC 1371 ($M_\text{SMBH}$) already exist. From the spiral arm pitch angle \citet{davis14} derived $M_\text{SMBH}=10^{6.17\pm0.5}$ M$_\odot$, close to Sgr A$^\star$, but from the well-known $M_{\text{SMBH}}-\sigma_\star$ relation \citep{woo13}, where $\sigma_\star$ is the velocity dispersion of the stellar bulge, \citet{cisternas13} estimated $M_{\text{SMBH}}\sim10^{7.4}$ M$_\odot$, an order of magnitude higher than Sgr A$^\star$. They also provided the bolometric luminosity ($L_{\text{bol}}\sim8\times10^{41}$ erg s$^{-1}$), from which we estimated the accretion rate ($\dot{m}$) onto the SMBH in NGC 1371 via
\begin{equation}
    \dot{m}=1.59\times10^{-26}\frac{L_\text{bol}}{\epsilon c^2}
,\end{equation}
where $\epsilon=0.1$ is the typical efficiency for the conversion of accreted mass into radiant energy (e.g., \citealt{heckman14}), $c$ is the speed of light and $1.59\times10^{-26}$ is the conversion factor from g s$^{-1}$ to M$_\odot$ yr$^{-1}$. This gives $\dot{m}\sim1.4\times10^{-4}$ M$_\odot$ yr$^{-1}$.
\\\indent In conclusion, we do not find conclusive evidence that the bubbles in NGC 1371 are extragalactic analogous to the FEBs. They share a similar medium temperature and size, but there are some differences in the mass and activity of the SMBH. Nevertheless, this does not exclude that they are the same class of object.

\subsection{The origin of the bubbles}
Bipolar nuclear bubbles are typically produced by AGN via jet-ISM interaction (see, e.g., \citealt{morganti98,ulvestad99,owen00,hota06,harrison15,muk18,li22,yang22,tanner22,konda23,zeng23}) or by a burst of nuclear star formation (e.g. \citealt{baum93,su10,crocker11,crocker15,yang18,bland19,ponti19,meliani24,thompson24}). Determining whether they originate from AGN activity or star-formation bursts is not straightforward for NGC 1371, given its low SFR and the uncertainty regarding their physical properties. From Fig. \ref{fig:index} we see that the bubbles have structures in the spectral index distribution. The inner part of the bubbles is dominated by the steep synchrotron emission from old, cooler electrons ($\alpha\ll-0.7$; e.g., \citealt{harwood13}), while on the edges the emitting medium is likely warmer, as the spectrum is flatter. Although the spectral index in the bubbles is inconsistent with the values for star-forming activity ($-0.8<\alpha_{\text{SFR}}<-0.6$), as mentioned in Sect. \ref{sec:agn} our analysis has caveats that need to be taken into account. Thus, without a proper correction for the contamination by thermal emission and without an extension of the frequency range, not possible with the current data, we cannot rule out that the bubbles are the remnant of a past burst of star formation in the stellar bar.
\\\indent Recent pc-resolution hydrodynamical simulations of the interaction between a $P_\text{jet}<10^{45}$ erg s$^{-1}$ jet and the surrounding medium have shown that kpc-scale bipolar bubbles, different from the radio lobes of the jet, are a natural consequence of the jet propagating within the gaseous disk at inclinations $>45^\circ$ with respect to the rotational axis of the galaxy \citep{muk18,tanner22}.
\\\indent With the $P_\text{jet}-L_{\text{1.4 GHz}}$ relation from \citet{heckman14}
\begin{equation}
    \text{Log}\left(\frac{P_\text{jet}}{\text{W}}\right)=4\times10^{35}(f_W)^\frac{3}{2}\text{Log}\left(\frac{L_{\text{1.4 GHz}}}{10^{25}\text{W Hz$^{-1}$}}\right)^{0.86}
,\end{equation}
originally found by \citet{willott99} (see also \citealt{cavagnolo10,daly12}), we estimated the jet power of the NGC 1371 AGN. Using $L^{\text{jet}}_{\text{1.4 GHz}}=(9.75\pm0.04)\times10^{20}$ W Hz$^{-1}$ derived in Sect. \ref{sec:radio} we found $P_\text{jet}\sim10^{39-41}$ erg s$^{-1}$ depending on the normalisation factor $f_W$ which is observationally restricted to be $1<f_W<20$ \citep{blundell00,hardcastle07}.
\\\indent Thus, the simulation by \citet{tanner22} is especially indicative as they simulate a jet with powers as low as $10^{41}$ erg s$^{-1}$, comparable with $P_\text{jet}$ estimated for NGC 1371. At such low luminosities, the jet inclination is the main driver of the bubble properties, including their temperature, composition, and pressure. They found that most of the outflow is expected to be ionised ($T>5000$ K), while the warm ($100\text{ K}<T<5000\text{ K}$) and cold ($T<100$ K) medium provide $<1\%$ of the total outflow mass. Unfortunately, no predictions are provided about the impact on the SFR.
\\\indent Other simulations of tilted jets predict that these bubbles exhibit a pronounced S-shaped limb brightening \citep{cecil00,wilson01,krause07} that is observed, for example, in M106 \citep{zeng23}. By carefully looking at the residual image of the continuum modelling, presented in the right-hand panel of Fig. \ref{fig:pybdsf}, we found tentative evidence for this limb brightening also in the NGC 1371 bubbles, as highlighted by the black dashed contours. Therefore, based on the aforementioned considerations, we argue that the bubbles in NGC 1371 could be originating from the impact of a $P_\text{jet}<10^{41}$ erg s$^{-1}$ jet propagating at angles $>45^\circ$ with respect to the rotational axis of the disk.
\\\indent However, the interaction of a more powerful jet in a previous cycle of activity or a burst in star formation from the rapid depletion of gas in the stellar bar might also have produced the bubbles. Further work on the age of the jet and the bubbles, and on the star formation history of NGC 1371 should help clarify the nature of the bubble.

\begin{figure*}
    \centering
    \includegraphics[width=\hsize]{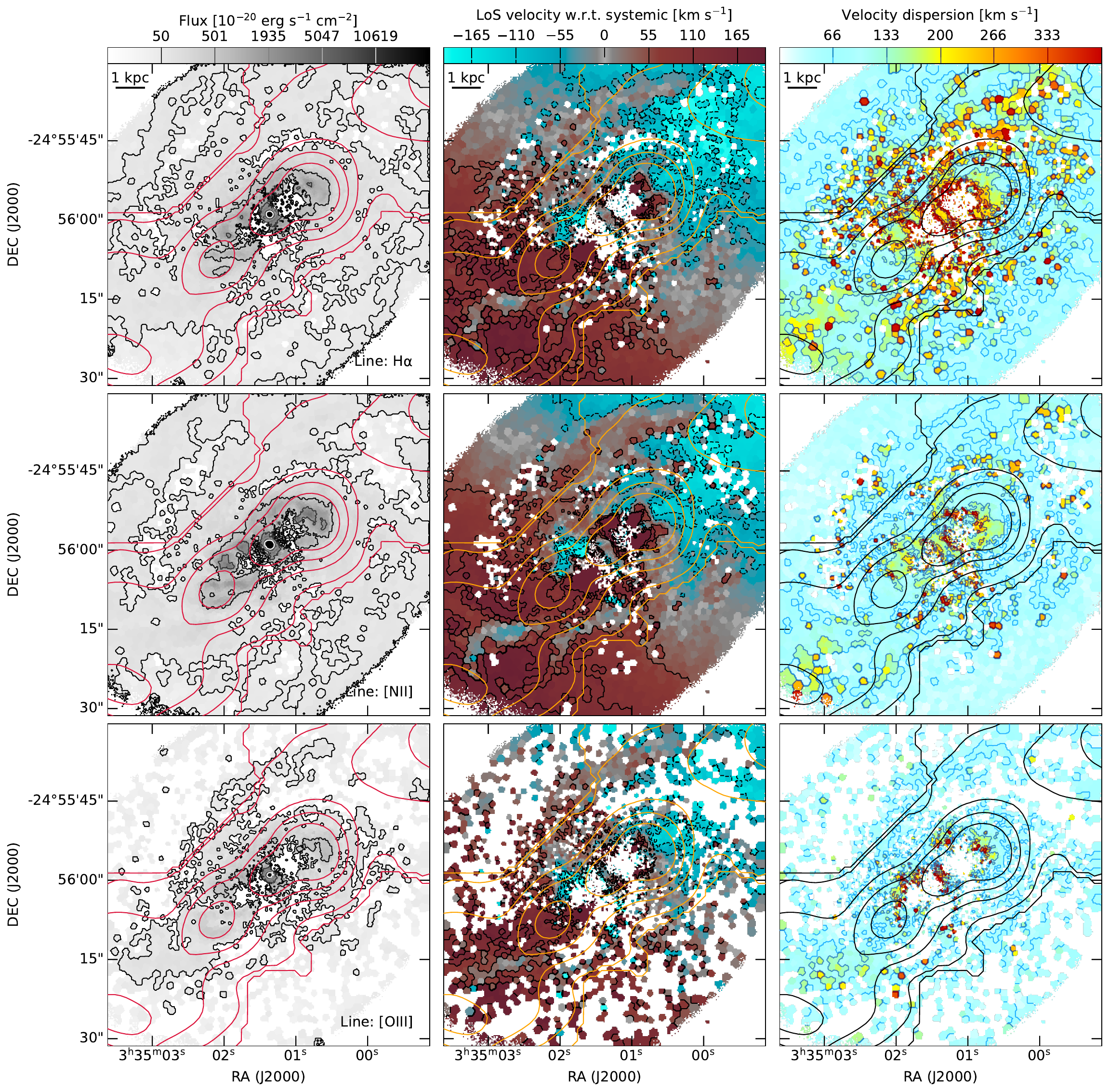}
    \caption{Ionised gas moment maps. \textit{Left columns}: intensity map in grey-scale from faint (grey) to bright (black) of the H$\alpha$, [NII] and [OIII] lines (from top to bottom). The contour levels for the intensity map are reported on the colorbar at the top. The red contours refer to the radio continuum (levels: 0.08, 0.31, 0.80, 1.68 mJy beam$^{-1}$). \textit{Central columns}: peak-velocity field with respect to the systemic velocity (1456.4 km s$^{-1}$). Dashed contours refer to the approaching side, while solid lines correspond to receding velocities. The radio continuum is overlaid with orange contours. \textit{Right panels}: Velocity dispersion map of the aforementioned lines. The contour levels are reported on the colorbar. The black contours refer to the radio continuum. In all the top panels the 1-kpc reference scale is shown in the top left.}
    \label{fig:ion}
\end{figure*}

\begin{figure*}
    \centering
    \includegraphics[width=\hsize]{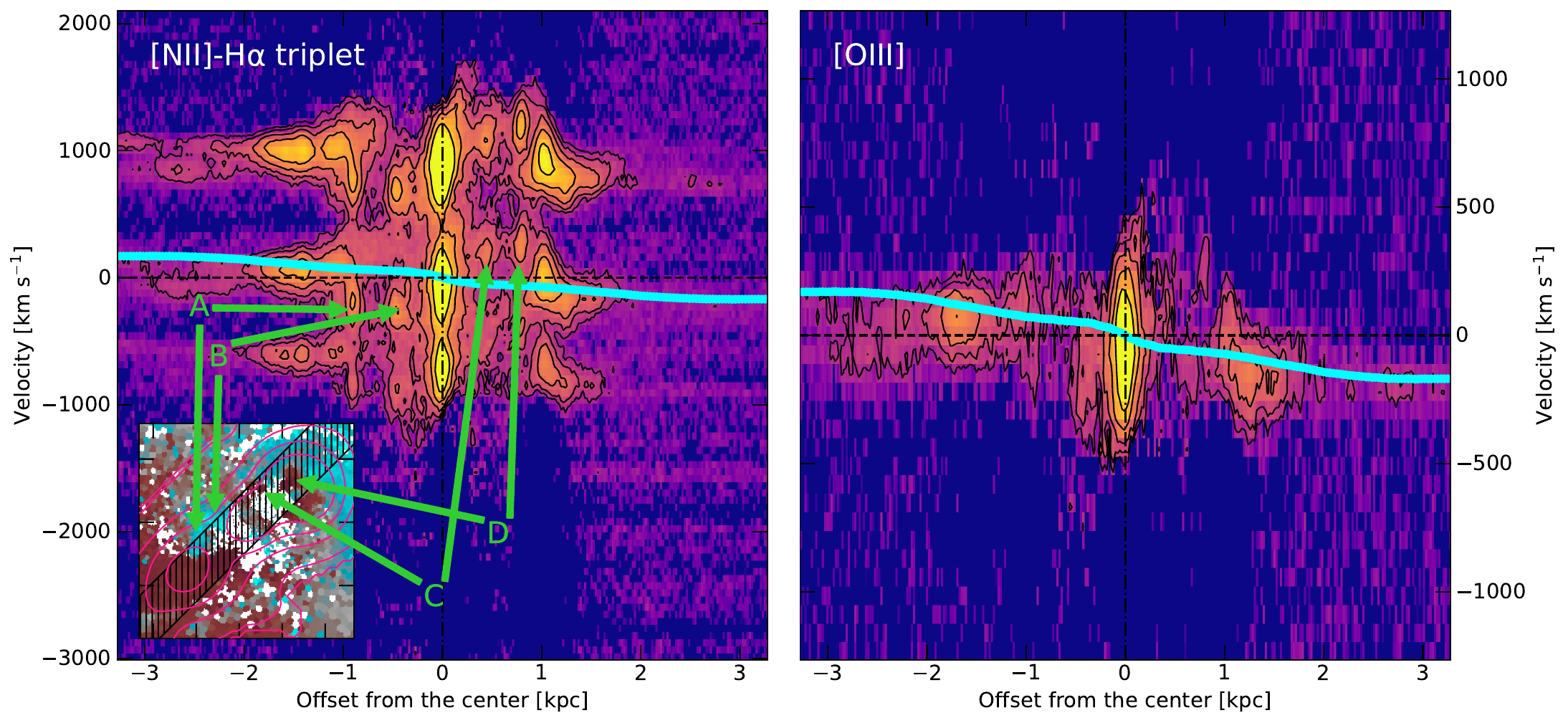}
    \caption{Position-velocity diagrams of the ionised gas. The left panel shows the $pv$ diagram along the major axis ($\theta=135^\circ$) of the [NII]-H$\alpha$ triplet, whereas the right-side panel shows the $pv$ diagram of the [OIII] line. Contours denote the (2, 4, 8, 16, 21)$\sigma$ level. The dashed black horizontal line corresponds to the H$\alpha$ and  [OIII] systemic velocity. The cyan line corresponds to the stellar rotation curve. The inset shows the velocity field of H$\alpha$, overlaid with the radio continuum (pink contours) and the slice from which the $pv$ is extracted (black hatched area). The arrows indicate kinematically anomalous features.}
    \label{fig:ionpv}
\end{figure*}

\subsection{AGN feedback}\label{sec:muse}
In the following we use multiwavelength data to learn about the feedback from the peculiar jet-ISM interaction occurring in this galaxy. Indeed, NGC 1371 is an outlier of the MHONGOOSE sample, being the only target significantly below the star-forming main sequence. It is also the only one where the presence of a kpc-scale jet is confirmed. Thus, it is natural to consider whether the presence of the central AGN is related to the low SFR of this galaxy.
\\\indent The jet and the base of the radio bubbles are (in projection) coincident with the $\sim5$-kpc wide H{\sc i} hole visible in the moment-0 map of Fig. \ref{fig:maps}. As we cannot use H{\sc i} to trace the state of the ISM in the inner 5 kpc, we require another gas tracer to study the feedback from the AGN. A tracer of the jet-ISM interaction are the optical emission lines of the ionised gas (e.g., \citealt{alatalo14,santoro15a,santoro15b,mahony16,venturi21,venturi23,maccagni21,tamhane23}). The characterisation of the ionised gas in the inner 5 kpc of the galaxy is achieved by extracting the emission line properties from the MUSE data presented in Sect. \ref{sec:anc}. Extracting the ionised gas kinematics involves modelling and subtracting the stellar continuum and fitting a set of Gaussian profiles to the residual cube. This can be done in one go with the \texttt{nGIST} pipeline \citep{ngist1,ngist2}, an extension of \texttt{GIST}\footnote{\url{http://ascl.net/1907.025}}, originally written by \citet{gist}. \texttt{nGIST} applies a preliminary SNR cut on the MUSE cube and further improves the SNR of the data via Voronoi binning \citep{voronoi1,voronoi2}. It then models the stellar continuum and gas kinematics using the penalised Pixel-Fitting Method (\texttt{pPXF}; \citealt{ppxf1,ppxf2,ppxf3}). The output of \texttt{nGIST} includes intensity, velocity, and dispersion maps of the stars and of the emission lines.
\\\indent We ran \texttt{nGIST} in wide-field mode with adaptive optics, because this is the observing mode that was used for the MUSE observations, and over the 4750-7100\AA{} range. This range avoids the noisiest regions of the spectrum and includes the diagnostic lines H$\beta_{\lambda4861.35}$, [OIII]$_{\lambda4958.91,\lambda5006.84}$, [NII]$_{\lambda6548.05,\lambda6583.45}$, H$\alpha_{\lambda6562.79}$, and [SII]$_{\lambda6716.44,\lambda6730.82}$ \citep{baldwin81,kewley01,kauffmann03,kewley06}. By visually inspecting the cube, we found that [NII] is the brightest line, which already suggests that the gas is not primarily excited by photoionisation from a starburst. Therefore, to maximise the information retrievable from the cube, we let \texttt{nGIST} to Voronoi bin the cube to achieve a minimum SNR of 50 per bin in the 6550-6650\AA{} range. For the stellar continuum modelling, we used the E-MILES stellar population synthesis templates from \citet{vazdekis16}, normalised to achieve a light-weighted best fit. The templates were fitted to the data with a set of 5-degree multiplicative and 11-degree additive Legendre polynomials\footnote{We tested different combinations of lower- and higher-degree polynomials and we found that this combination provides the best fit.}. For fitting the gas emission lines, we tied the kinematic parameters of the emission lines into three groups \citep{phangsmuse}: Balmer (H$\beta$ and H$\alpha$), low ionisation ([OI]$_{\lambda6300,\lambda6364}$, [NI]$_{\lambda5197,\lambda5200}$, [NII], [NII]$_{\lambda5754}$, and [SII]), and high-ionisation ([HeI]$_{\lambda5875}$, [OIII], and [SIII]$_{\lambda6312}$) lines. This classification is because the main ionisation mechanism differs for each group: AGN and star formation for Balmer lines, star formation for low-ionisation lines, and AGN for high-ionisation lines \citep{heckman80,baldwin81}.
\\\indent In Fig. \ref{fig:spec}, we present the integrated spectrum of the inner 5 kpc of the galaxy. We also show the best-fit stellar continuum, the emission lines from the ionised gas, and the residuals, demonstrating good agreement between the fit and the data. High residuals are found in the 5775-6010\AA{} range, i.e., the band affected by the adaptive optics laser. The flux in these channels is set by \texttt{nGIST} to the median value of the spectrum and does not enter the fit. There are also two bright spikes around 6300\AA{}. These are not real emission lines, but faulty channels.
\\\indent In Fig. \ref{fig:ion}, we present the intensity, line-of-sight velocity, and velocity dispersion\footnote{The velocity dispersion has been corrected for the instrumental broadening: $\sigma_0=\sqrt{\sigma^2-\sigma^2_\text{ins}}$.} maps for the three line groups, overlaid with the radio continuum contours. Bins with a poor fit are blanked. These were defined by $\delta v>\sigma_\text{ins}$ and $\delta\sigma>\sigma_\text{ins}$, where $\delta v$ and $\delta\sigma$ are the error in the velocity and dispersion of the line and $\sigma_\text{ins}$ is the instrumental broadening. $\sigma_\text{ins}$ depends on the considered wavelength: for [OIII] is 64 km s$^{-1}$, while for H$\alpha$ and [NII] is 49 km s$^{-1}$. These selection criteria were supported by the visual inspection of the spectra. The maps show that the brightest emission from the ionised gas originates within the inner hotspots of the jet. Within the innermost 1.5 kpc, the gas is not following the disc rotation and its emission is broader.
\\\indent The presence of non-rotating gas is also evident from the channel maps in a 800 km s$^{-1}$ range around H$\alpha$, provided in Fig. \ref{fig:gaschanmap}. This velocity range is sufficient to cover the emission from H$\alpha$ and prevent significant contamination from the wings of the [NII] doublet, as indicated by the dispersion map in the right-side panels of Fig. \ref{fig:ion}. The maps further corroborate the spatial coincidence between H$\alpha$ and the bright radio continuum tracing the inner regions of the jets. This suggests that within 1.5 kpc from the centre the jet is impacting ISM, affecting its kinematics and physical state.
\\\indent The output of the \texttt{nGIST} pipeline includes the stellar kinematics map obtained from the fit to the stellar absorption lines. We use the map to derive the rotation curve of the stellar disc. This has been computed with the \texttt{2DFIT} task of \texttt{3D-Barolo}. We fitted the rotation velocity and position angle of 100 rings, each 0.3$''$ wide. We fixed the inclination of the rings to 46$^\circ$, i.e. the optical inclination of the galaxy as listed in \citet{mhongoose2}, the radial velocities to 0 km s$^{-1}$, and the systemic velocity also to 0 km s$^{-1}$, as the kinematics map is already corrected for it. We use this information to compare with the ionised gas kinematics. The 7$''$-wide $pv$ diagram extracted along the major axis, given in Fig. \ref{fig:ionpv}, shows this comparison. The identification of the aforementioned kinematically anomalous gas, indicated by the arrows in Fig. \ref{fig:ionpv}, suggests again that probably a jet-induced outflow is present in the nuclear region.
\\\indent To further explore the jet-ISM interaction in NGC 1371 we investigated the impact of the radio jet on the molecular gas. Indeed, this phenomenon has been well studied with CO observations (e.g., \citealt{alatalo11,cicone14,morganti15,morganti21,oosterloo17,oosterloo19,fluetsch19,ruffa20,ruffa22,maccagni21,audibert23}). We visually inspected the low-resolution ACA cube (beam: $\Omega=14''\times14''$; channel separation: $\Delta v=4.4$ km s$^{-1}$) presented in Sect. \ref{sec:anc} resulting in no detection. Searching for emission with \texttt{SoFiA-2} also provides no reliable sources at the $3\sigma$ level. After converting the brightness temperature noise value ($\sigma_{T_{\rm b}}=3$ mK) into Jy \citep{meyer17}
\begin{equation}
    \frac{\sigma_{\rm S}}{\text{Jy}}=4.8\times10^{12}\frac{k_{\rm b}\nu_{\text{obs}}^2\Omega\sigma_{T_{\rm b}}}{c^2}
\end{equation}
where $k_{\rm b}$ is the Boltzmann constant, $\Omega$ is the beam area in arcsec$^{2}$ and $4.8\times10^{12}$ accounts for the steradians-to-arcsec$^2$ and Jy-to-erg s$^{-1}$ conversions, we calculated an upper limit on the molecular mass (M$_{\text{H$_2$}}$) using the following equation:
\begin{equation}
    M_\text{H$_2$}=L_{\rm CO}\cdot\alpha_{\rm CO}
\end{equation}
where $\alpha_{\rm CO}=4.36$ M$_\odot$ is a standard CO-to-H$_\text{2}$ conversion factor \citep{tacconi13,bolatto13,audibert22,audibert23} and
\begin{equation}
    \frac{L_{\rm CO}}{\text{K km s$^{-1}$ pc$^{-2}$}}=3.25\times10^7\left(\frac{\sigma_{\rm S}}{\text{Jy}}\right)\left(\frac{\Delta v}{\text{km s$^{-1}$}}\right)\left(\frac{D_{\rm L}}{\text{Mpc}}\right)^2\left(\frac{\nu_{\rm obs}}{\text{Hz}}\right)^{-2}
\end{equation}
is the CO luminosity \citep{solomon05}. The estimated upper limit in the inner 5-kpc of the galaxy (i.e., within the H{\sc i} hole) is $M^\text{hole}_\text{H$_2$}<2\times10^5\text{ M$_\odot$}$. This value is three orders of magnitude lower than the typical molecular mass detected in circumnuclear disks of AGNs (e.g., \citealt{maccagni18,ruffa19a,combes19}) and about two dex lower if we compare with the upper limits for non-detected disks \citep{tadhunter24}. This may indicate a lack of molecular gas or that it is just warmer ($T\sim$tens of K, \citealt{oosterloo17}). In the latter scenario, the CO$_{2\to1}$ or CO$_{3\to2}$ emission lines are much brighter than CO$_{1\to0}$ (e.g. \citealt{alatalo11,oosterloo17,oosterloo19,ruffa22}). Observations using these lines would clarify the lack of detectable CO$_{1\to0}$.
\\\indent In conclusion, we found indications that in the inner 3 kpc of NGC 1371 the low-power jet is producing an outflow in the ionised phase of the gas. We found no detectable CO$_{1\to0}$ down to $2\times10^5\text{ M$_\odot$}$ within the inner 5 kpc, where also H{\sc i} is not detected up to column densities of 1.1 $\times10^{19}$ cm$^{-2}$. Why does only the hot phase of the gas survive in the central region?

\subsection{The absence of cold ISM}\label{sec:bar}
Jet-ISM coupling typically does not prevent the cooling of the gas and the survivability of neutral and molecular hydrogen. Evidence is the large amount of H{\sc i} (e.g., \citealt{morganti03,morganti16,morganti18,gupta06,chandola11,mahony13,gereb15a,maccagni17}) and CO (e.g. \citealt{alatalo11,cicone14,morganti15,morganti21,oosterloo17,oosterloo19,fluetsch19,ruffa20,ruffa22,maccagni21,audibert23}) observations of such interactions, as well as the support from theoretical works (e.g., \citealt{muk18,muk18b,tanner22}). Also, the effect of the jet is directional, while in NGC 1371 the non-detection of cold gas is ubiquitous. However, since the gas in the galaxy rotates, one possible way out is that in the inner 5 kpc of NGC 1371 the gas cooling time is higher than its orbital time. Assuming the gas corotates with the stellar disk on circular orbits, from the stellar rotation curve derived earlier we calculate the orbital period. This ranges from $\sim10$ to $\sim70$ Myr, which implies cooling times of similar amplitudes. This is not an unreasonable range, given the many physical parameters affecting the cooling of the gas, such as its density, metallicity, and temperature (e.g., \citealt{valentini15,kanjilal21,dutta24}).
\\\indent Another possibility is that the gas is blown out by a wind. Based on observations of 94 AGN with detected sub-pc to kpc winds, \citet{fiore17} argued that AGN winds are very efficient in removing cold gas from the nuclear region of galaxies. At $L_\text{bol}\sim10^{45}$ erg s$^{-1}$ the wind reaches velocities of $\sim700$ km s$^{-1}$ and the molecular mass outflow is $\sim200$ M$_\odot$ yr$^{-1}$, while at $L_\text{bol}\sim10^{42}$ erg s$^{-1}$ the wind speed is $\sim200$ km s$^{-1}$ and the outflow reaches $\sim1$ M$_\odot$ yr$^{-1}$. We recall that the estimated bolometric luminosity for NGC 1371 is $L_\text{bol}\sim10^{42}$ erg s$^{-1}$ \citep{cisternas13}, implying wind speeds of $\sim200$ km s$^{-1}$ and an outflow rate of $\sim1$ M$_\odot$ yr$^{-1}$. This is reasonable for NGC 1371, as the velocity dispersion of the ionised gas, which could be used to trace also the molecular wind speed \citep{fiore17}, is $\sim200$ km s$^{-1}$ along the radio jet (see the top-right panel of Fig. \ref{fig:ion}). 
\\\indent Another process capable of removing gas from the central region of galaxies is bar quenching \citep{khoperskov18,george19}. Stellar bars are commonly found in the inner regions of disk galaxies (e.g., \citealt{menendez07,hernandez07,marinova07} and for our Galaxy \citealt{binney91,blitz91,weiland94}) and there is growing observational \citep{masters10,cheung13,gavazzi15,james16,kim17,wang20,scaloni24} and theoretical (e.g., \citealt{carles16,spinoso17}) evidence, even in our Galaxy \citep{haywood16,haywood18}, that bars favour the rapid depletion of gas and the inside-out quenching of disk galaxies. Based on near-infrared imaging with the Infrared Array Camera (IRAC; \citealt{irac}), \citet{cisternas13} classified NGC 1371 as a barred galaxy (see also \citealt{kim15}). Therefore, it is plausible that in this galaxy the lack of cold ISM in the inner 5 kpc results from the rapid depletion of gas from the bar.
\\\indent Ultimately, we found that the low-power, inclined jet propagating in an under-dense ISM could explain the inside-out quenching of NGC 1371, although the stellar bar might play a significant role as well. Future hydrodynamical simulations of tilted jets and future observations of a sample of galaxies sharing similar properties (stellar bar, H{\sc i} hole, low-power tilted jet) should be able to favour one of the possible scenario we proposed to explain the lack of cold gas in the inner 5 kpc: the ISM is not cold but warm, hence should be traced via CO$_{2\to1}$ or CO$_{3\to2}$ emission lines; the ISM has been mostly blown out by the AGN wind or it has been depleted by the stellar bar.

\begin{figure*}
    \centering
    \includegraphics[width=\hsize]{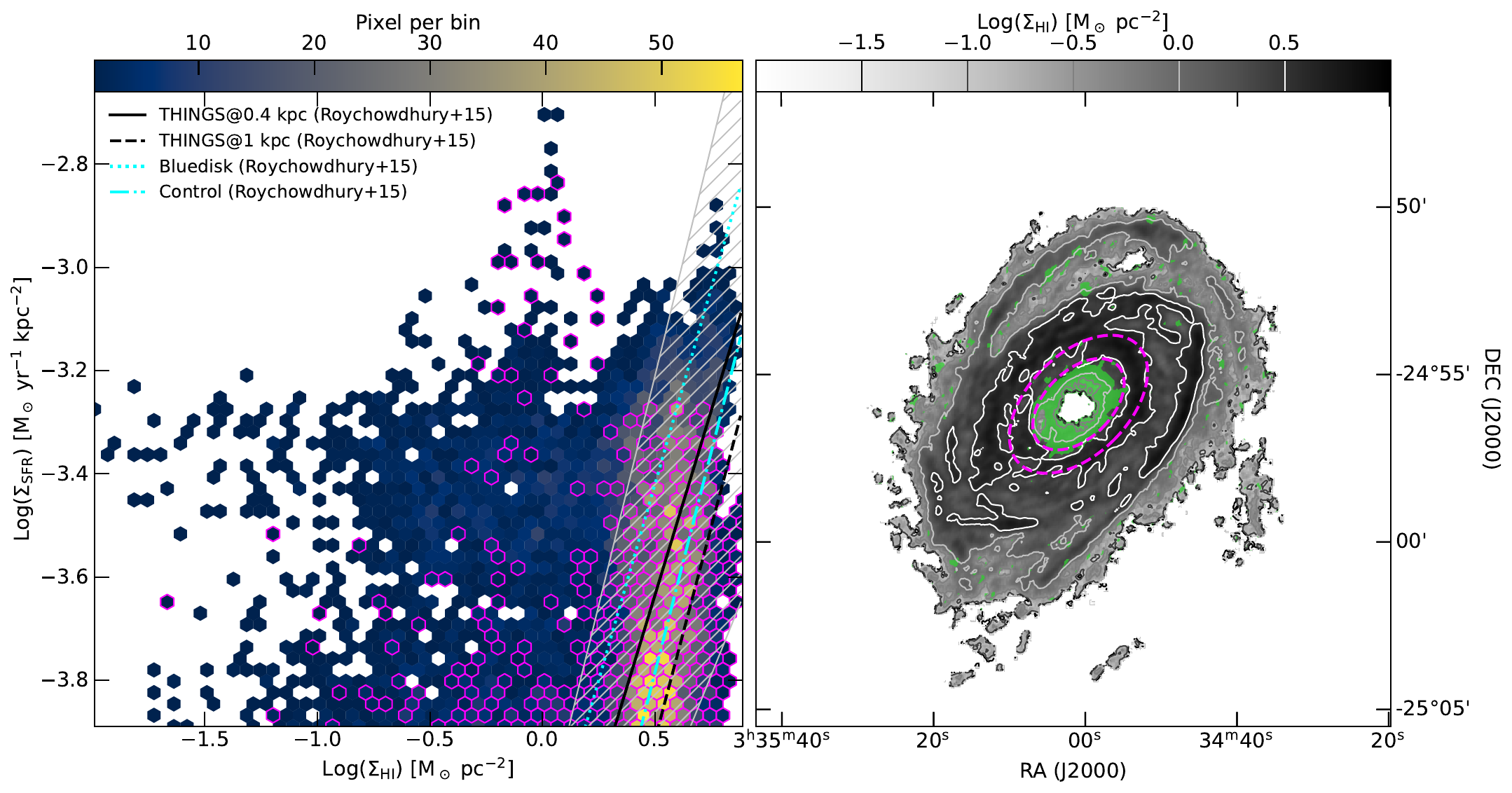}
    \caption{Spatially resolved Kennicutt-Schimdt law. \textit{Left panel}: density plot of $\Sigma_{\text{SFR}}$ as a function of $\Sigma_{\text{H{\sc i}}}$. The two-dimensional histogram is made of $49\times49$ hexagonal bins, each containing at least 1 pixel. The coloured lines indicate the relations found by \citet{roy15} for H{\sc i} dominated regions in spiral galaxies: solid and dashed black curves refer to the relation found for THINGS \citep{things} galaxies studied at 0.4 kpc resolution (solid line) and 1 kpc resolution (dashed line), whereas the dotted cyan line indicates the relation found for the Bluedisk sample \citep{bluedisk} and the dash-dotted line refers to the control sample used by \citet{roy15}. The silver-hatched area indicates the $\pm5\sigma$ scatter of the relation relative to the THINGS galaxies at 0.4 kpc resolution. Bins with magenta edges correspond to pixels within the dashed magenta elliptical annulus in the right-side panel. \textit{Right panel}: $\Sigma_{\text{H{\sc i}}}$ map. Contour levels are reported on the colorbar. The green regions indicate the location of the pixel which are in the hatched area in the left-side panel. The dashed magenta elliptical annulus provides the approximate location of a star forming ring observed outside the H{\sc i} hole.}
    \label{fig:ksmap}
\end{figure*}

\subsection{Star formation in the H{\sc i} disc}\label{sec:kslaw}
Having established the dual relevance of the stellar bar and the low-luminosity AGN in perturbing the gas in the inner 5 kpc of the galaxy, we want to understand how far out in the gaseous disc the influence of these two processes extends. This requires studying whether the SFR is consistent with the gaseous content of the disk. The relation between the gas density of the ISM and the SFR is given by the Kennicutt-Schmidt law \citep{schmidt59,kennicutt98,kennicutt12,reyes19}. This relation is well known and very tight for the molecular phase of the gas (e.g., \citealt{bigiel08,marasco12,leroy13,reyes19}), yet the precise correlation between $\Sigma_{\text{SFR}}$ and the H{\sc i} surface density ($\Sigma_{\text{H{\sc i}}}$) remains a matter of debate (e.g., \citealt{kennicutt07,leroy08,bigiel10,roy15,yim16,bacchini19,eibensteiner24}).
\\\indent The computation of the $\Sigma_{\text{SFR}}$ maps is provided in Sect. \ref{sec:radio}. We derived $\Sigma_{\text{H{\sc i}}}$ from the inclination and primary beam-corrected moment 0 map ($I_\text{H{\sc i}}$). First, we converted $I_\text{H{\sc i}}$ into column density ($N_{\text{H{\sc i}}}$; \citealt{meyer17}):
\begin{equation}\label{eq:colden}
    \frac{N_{\text{H{\sc i}}}}{\text{atoms cm}^{-2}}=1.25\times10^{24}\left(\frac{I_\text{H{\sc i}}}{\text{Jy km s}^{-1}}\right)\left(\frac{\Omega}{\text{arcsec}^2}\right)^{-1}
,\end{equation}
where $\Omega$ is the beam area.
Then, we computed $\Sigma_{\text{H{\sc i}}}$ using
\begin{equation}\label{eq:hi}
    \frac{\Sigma_{\text{H{\sc i}}}}{\text{M$\odot$ pc$^{-2}$}}=7.95\times10^{-19}\left(\frac{N_{\text{H{\sc i}}}}{\text{atoms cm}^{-2}}\right)
,\end{equation}
where $7.95\times10^{-19}$ accounts for the cm-to-parsec and atoms-to-M$_\odot$ conversions. In the left-hand panel of Fig. \ref{fig:ksmap} we show the spatially resolved $\Sigma_{\text{SFR}}$ as a function of $\Sigma_{\text{H{\sc i}}}$. We also plot the relations found by \citet{roy15} for H{\sc i}-dominated regions in spiral galaxies from the THINGS \citep{things}, the Bluedisk \citep{bluedisk} and a control sample. The majority of the points are consistent with these relations, thus, most of the H{\sc i} disc follows the expected Kennicutt-Schmidt law.
\\\indent In the right-hand panel of Fig. \ref{fig:ksmap}, we instead plot $\Sigma_{\text{H{\sc i}}}$ overlaid with the locations of the pixels within the hatched area. Examining their spatial distribution, we found that most of the outliers are located around the H{\sc i} hole. The reason for their displacement from the Kennicutt-Schmidt law is possibly that in this region of the disc $\Sigma_{\text{H{\sc i}}}$ is decreasing towards the center, whereas $\Sigma_{\text{SFR}}$ is not, as shown in Fig. \ref{fig:sfrhi}. One explanation could be that the H{\sc i} is turning molecular. Indeed, H{\sc i} holes in the centres of spiral galaxies are known to result from the conversion of H{\sc i} into H$_2$ \citep{leroy08,bigiel08}. However, based on the very low upper limit on the molecular mass, derived at the beginning of this section ($M_{\text{H$_2$}}<5\times10^6$ M$_\odot$), we think this is unlikely. It is more probable that $\Sigma_{\text{H{\sc i}}}$ is decreasing due to the dual effect of the low-power AGN and the bar quenching, as explained in Sect. \ref{sec:muse} and in Sect. \ref{sec:bar}.
\\\indent From the multiwavelength image in Fig. \ref{fig:overlay} and the SFR map of Fig. \ref{fig:sfr} it is evident a ring of star formation just outside the H{\sc i} hole. In the right-side panel of Fig. \ref{fig:ksmap} we show its approximate location with the dashed magenta elliptical annulus. The spatially resolved KS law for the pixels in the annulus is reported in left-side panel of Fig. \ref{fig:ksmap} with the bins having magenta edges. These bins are mostly concentrated around the relations for H{\sc i}-dominated regions, meaning they follow the KS law. Therefore, the star forming ring may indicates the radii at which the inside-out quenching stops, and the gas is forming stars as expected in H{\sc i} dominated regions.
\\\indent In conclusion, we found that up to $\sim10$ kpc from the centre, the gas might still be affected by the processes that are removing or heating the ISM within the H{\sc i} hole, i.e., AGN winds and/or the stellar bar. At larger radii, however, the disc is forming stars as expected from the Kennicutt-Schmidt relation of H{\sc i}-dominated regions in spiral galaxies. Yet, the inner quenching of star formation appears enough to prevent the formation of the $\sim1.75$ M$_\odot$ yr$^{-1}$ needed to bring the galaxy back on the star forming main sequence.

\subsection{The impact of the environment}\label{sec:env}
So far we have considered only the internal processes affecting the star formation and the state of the gas in NGC 1371. It is known that also external processes tidal interaction and gas stripping are able to impact the star formation of galaxies from both theoretical \citep{hirschmann14,hatfield17} predictions and empirical \citep{peng15,knobel15} evidence. As mentioned in Sect. \ref{sec:intro}, NGC 1371 is a member of the Eridanus group \citep{brough06}, therefore it is worth investigating if there is any environmental effect acting on this galaxy. In Sect. \ref{sec:hi} we already excluded a possible recent or ongoing interaction between NGC 1371 and the newly-discovered companion MKT 033442-245148. The H{\sc i} disk is also not showing strong tidal features, like filaments and tails, but it is warped in the outermost regions. Thus, it can still be that the H{\sc i} disk of NGC 1371 is being perturbed by the presence of the nearby companions.
\\\indent The Eridanus group comprises at least 35 galaxies \citep{for23}. Using WALLABY pre-pilot data, \citet{for23} studied the H{\sc i} properties of this group, finding that all the galaxies show a distorted H{\sc i} morphology. They suggests that group interactions are occurring. The most massive neighbour of NGC 1371 is NGC 1385 ($\text{Log}M_*=10.13$, $\text{Log}M_{\text{H{\sc i}}}=9.34$; \citealt{for23}). Its distance ($D$) from NGC 1371 can be roughly estimated using
\begin{equation}
    \frac{D}{\text{Mpc}}=\sqrt{\left(\frac{dl}{\text{Mpc}}\right)^2+\left(\frac{d\alpha}{\text{Mpc}}\right)^2+\left(\frac{d\delta}{\text{Mpc}}\right)^2}
,\end{equation}
where $dD$, $d\alpha$ and $d\delta$ are the distance and the projected right-ascension and declination difference, respectively. Using the distance provided by \citep{leroy19} and the fiducial position reported in NED for NGC 1385, we estimate $D\sim0.3$ Mpc. This distance is lower than the AMIGA isolation radius\footnote{The isolation radius for a galaxy of diameter $d_0$ with respect to a neighbour having diameter $0.25d_0<d_1<4d_0$ is equal to $20d_1$.} \citep{verdes05} of $\sim0.9$ Mpc, assuming an H{\sc i}-based diameter for NGC 1385 and NGC 1371 of $\sim46$ kpc and $\sim84$ kpc, respectively, and lower than the NGC 1371 virial radius $R_{200}\sim0.37$ Mpc \citep{mhongoose2}. From the above considerations, it is plausible that a weak interaction is ongoing between NGC 1371 and NGC 1385, but evident signatures of gas removal are not found. Therefore, we argue that the environment plays a minor role in the quenching of NGC 1371.

\section{Conclusions and future prospects}\label{sec:conc}
In this paper we presented the deepest 21-cm spectral line and 1.4 GHz continuum observation ever taken of the nearby spiral galaxy NGC 1371. This galaxy is below the star-forming main sequence and we aim to understand why is that. In the following we list our main findings:
\begin{enumerate}[i)]
    \item The radio continuum reveals a low-luminosity ($L^{\text{jet}}_{\text{1.4 GHz}}=(9.75\pm0.04)\times10^{20}$ W Hz$^{-1}$) jet extending up to 5 kpc (in projection) from the galactic centre;
    \item In the radio continuum we also detected $\sim10$-kpc wide bipolar nuclear bubbles, having total luminosity of $L^{\text{bubbles}}_{\text{1.4 GHz}}<4.6\times10^{20}$ W Hz$^{-1}$. Such structures have never been observed before in NGC 1371 and are rarely seen in other galaxies;
    \item We do not find compelling evidence that the bubbles in NGC 1371 are an extragalactic analogue of our Galaxy FEBs, yet we cannot exclude they are the same class of object;
    \item The presence of such bubbles and their S-shaped limb brightening suggests that the jet in NGC 1371 is likely emitted with an angle $>45^\circ$ with respect to the galaxy rotational axis;
    \item There are indications that the tilted jet is producing an ionised gas outflow in the innermost kpc (in projection).
    \item The H{\sc i} is mostly distributed in a regularly rotating disc and peculiar features includes a warped outer disc and a $\sim5$ kpc-wide H{\sc i} hole around the galactic centre.
    \item within the H{\sc i} hole the molecular gas traced by the CO$_{1\to0}$ transition is non-detected, allowing us to compute only an upper limit (M$_{\text{H$_2$}}<2\times10^5\text{ M$_\odot$}$);
    \item The lack of cold gas can be explained if the ISM affected by the AGN is warm, or if its cooling time is longer than its orbital time ($<70$ Myr), or if it has been blown out by AGN winds, or depleted by the stellar bar. We argue that one (or more) of these four scenarios is causing the inside-out quenching of NGC 1371;
    \item At larger distances, the gaseous disc is forming stars as expected from the Kennicutt-Schmidt relation of H{\sc i}-dominated regions in spiral galaxies;
    \item As far as the galaxy environment is concerned, we identified a new galaxy. MKT 033442-245148 ($\alpha=3^h34^m42.31^s$, $\delta=-24^d51^m47.52^s$, $cz=1647$ km s$^{-1}$) has an H{\sc i} mass of $M_{\rm H{\sc i}}=6.6\times10^6$ M$_\odot$ and we did not find strong evidence for a recent interaction with NGC 1371. Instead, from the isolation radius of NGC 1371 we found that the galaxy is probably weakly interacting with the neighbour NGC 1385, although we do not have enough element to quantify the impact on its quenching.
\end{enumerate}
Some aspects remain still unclear. In the future, deriving the star formation history of NGC 1371 and the age of its bubbles is fundamental to understand if the bubbles could be the remnant of a past star burst in the stellar bar. Observing NGC 1371 centre with higher excitation lines of CO ( CO$_{2\to1}$ and CO$_{3\to2}$) would be useful to confirm if our CO$_{1\to0}$ non-detection is due to the lack of molecular gas or a higher medium temperature. It would be useful also to further investigate 3D hydrodynamical simulations of a low-power ($P_\text{jet}<10^{41}$ erg s$^{-1}$) jet propagating at an angle $>45^\circ$ with respect to the galaxy rotational axis within a medium pre-processed by a stellar bar or an AGN wind. This will provide insights into why in NGC 1371 we are not detecting the neutral and molecular ISM in the inner 5 kpc of this galaxy. Moreover, such simulations allow to study the inside-quenching in this peculiar environment.

\begin{acknowledgements}
We thank the anonymous referee for the constructive comments.\\\indent 
This work has received funding from the European Research Council (ERC) under the European Union’s Horizon 2020 research and innovation programme (grant agreement No 882793 `MeerGas').\\\indent FMM acknowledges that (part of) the research activities described in this paper were carried out with contribution of the Next Generation EU funds within the National Recovery and Resilience Plan (PNRR), Mission 4 - Education and Research, Component 2 - From Research to Business (M4C2), Investment Line 3.1 - Strengthening and creation of Research Infrastructures, Project IR0000034 – “STILES - Strengthening the Italian Leadership in ELT and SKA.\\\indent This research has made use of the NASA/IPAC Extragalactic Database (NED), which is funded by the National Aeronautics and Space Administration and operated by the California Institute of Technology.\\\indent This paper makes use of the following ALMA data: ADS/JAO.ALMA\#2022.1.01314.S. ALMA is a partnership of ESO (representing its member states), NSF (USA) and NINS (Japan), together with NRC (Canada), NSTC and ASIAA (Taiwan), and KASI (Republic of Korea), in cooperation with the Republic of Chile. The Joint ALMA Observatory is operated by ESO, AUI/NRAO and NAOJ.\\\indent SV thanks E. di Teodoro for the insightful discussions about kinematical modelling of galaxy discs.
\end{acknowledgements}

\bibliographystyle{aa}
\bibliography{reference.bib}

\appendix
\onecolumn
\section{Ancillary plots}\label{app:barolo}

\begin{figure*}[!h]
    \centering
    \includegraphics[width=\hsize]{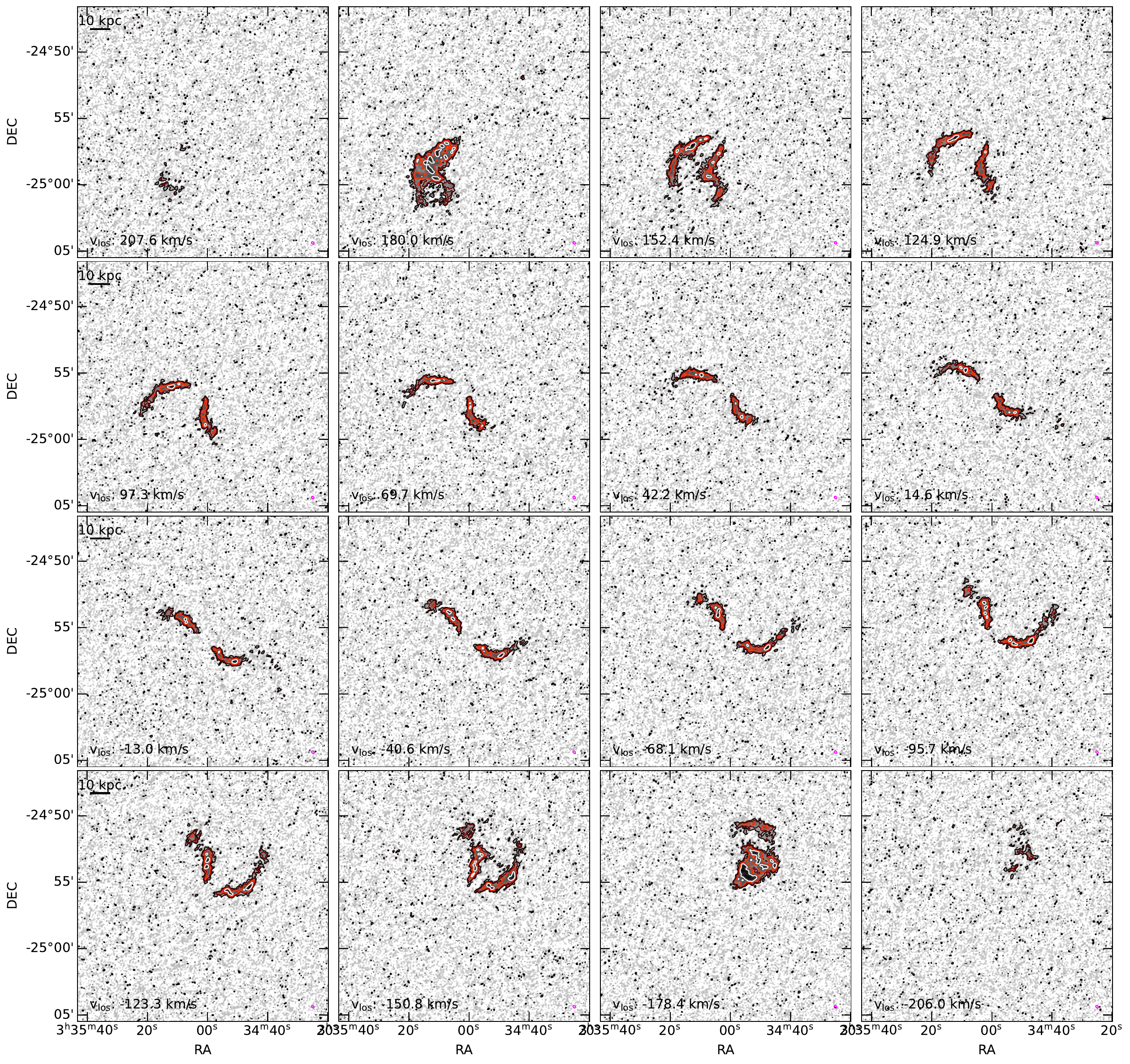}
    \caption{NGC 1371 H{\sc i} channel maps from 1250 km s$^{-1}$ to 1665 km s$^{-1}$ in steps of 28 km s$^{-1}$. The coloured contours correspond to column densities of (4.3, 8.6, 17.3, 34.5) $\times$ 10\textsuperscript{18} cm$^{-2}$, equivalent to (4, 8, 16, 32)$\sigma$. The dashed grey contours denote the column density of $-4.3\times10^{18}$ cm$^{-2}$. At the bottom, we report the line-of-sight velocity (with respect to the systemic velocity) in each channel and at the top left the 10-kpc reference scale. In the bottom-right corner the $13.7''\times9.6''$ beam is displayed.}
    \label{fig:chanmap}
\end{figure*}

\begin{figure}
    \centering
    \includegraphics[width=.83\hsize]{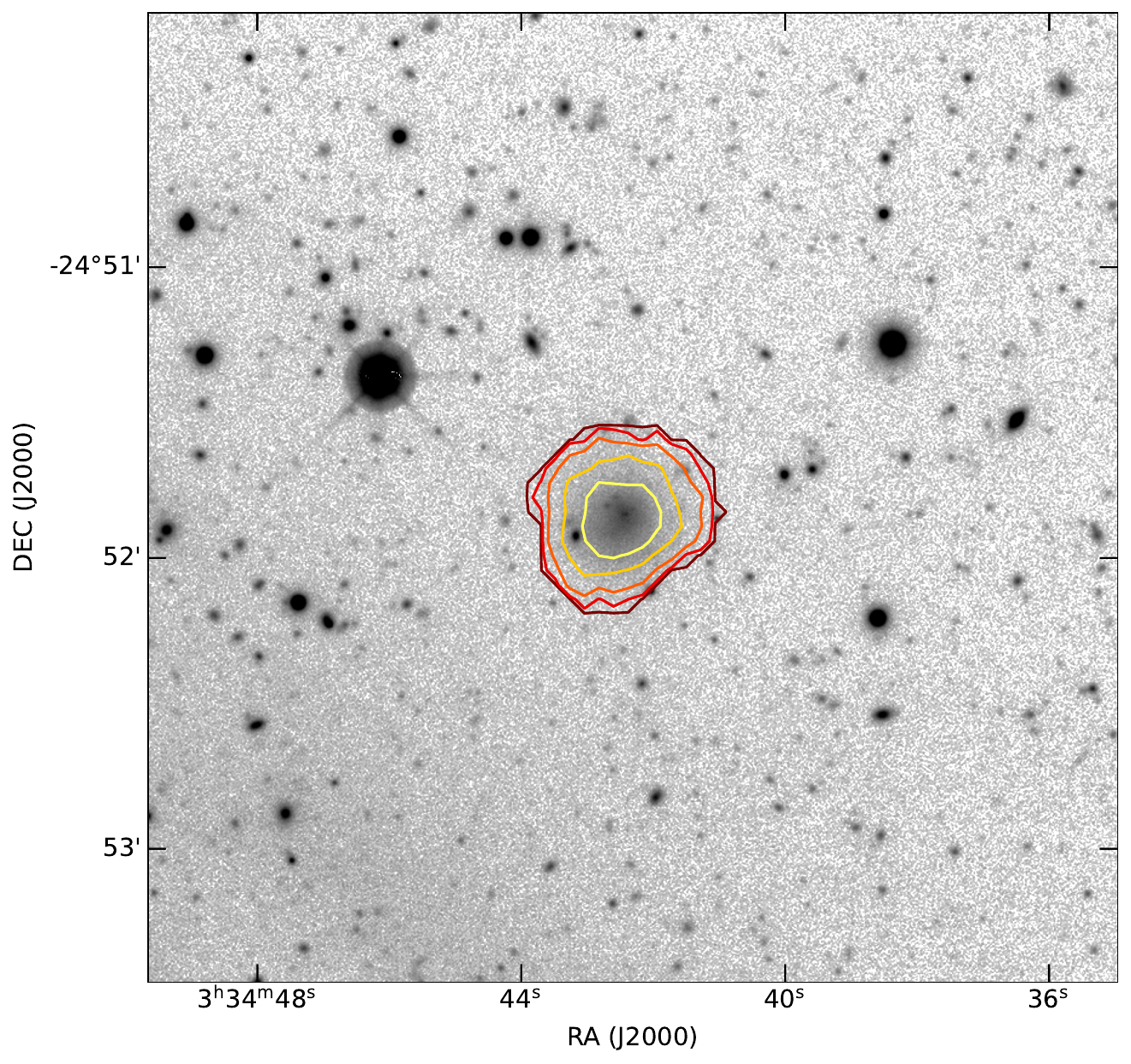}
    \caption{Combined $gzri$ DECaLS image of MKT 033442-245148 overlaid with the H{\sc i} contours denoting the (0.1, 0.6, 2.2, 5.8, 12.1)$\times10^{19}$ cm$^{-2}$ column density levels.}
    \label{fig:clump}
\end{figure}

\begin{figure*}
    \centering
    \includegraphics[width=\hsize]{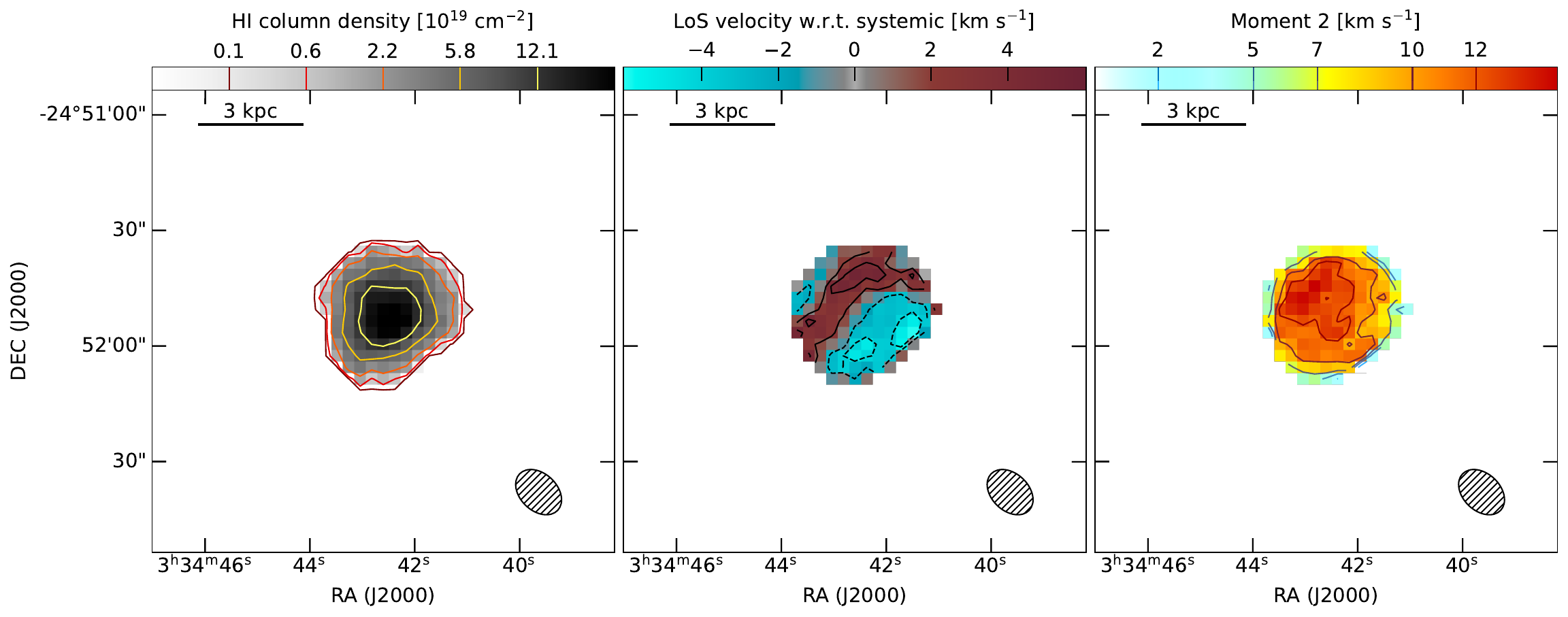}
    \caption{High-resolution H{\sc i} moment maps of MKT 033442-245148. \textit{Left panel}: primary beam-corrected H\,{\sc i} column-density map. \textit{Central panel}: Intensity-weighted mean velocity field with respect to the systemic velocity (1647 km s$^{-1}$). Dashed contours refer to the approaching side, while solid lines correspond to receding velocities. \textit{Right panel}: Second moment map. In all panels the contour levels are given on the colorbar, while in the bottom right we show the $13.7''\times9.6''$ beam and in the top left corner the 3-kpc reference scale.}
    \label{fig:clumpmaps}
\end{figure*}

\begin{figure*}[!h]
    \centering
    \includegraphics[width=.87\hsize]{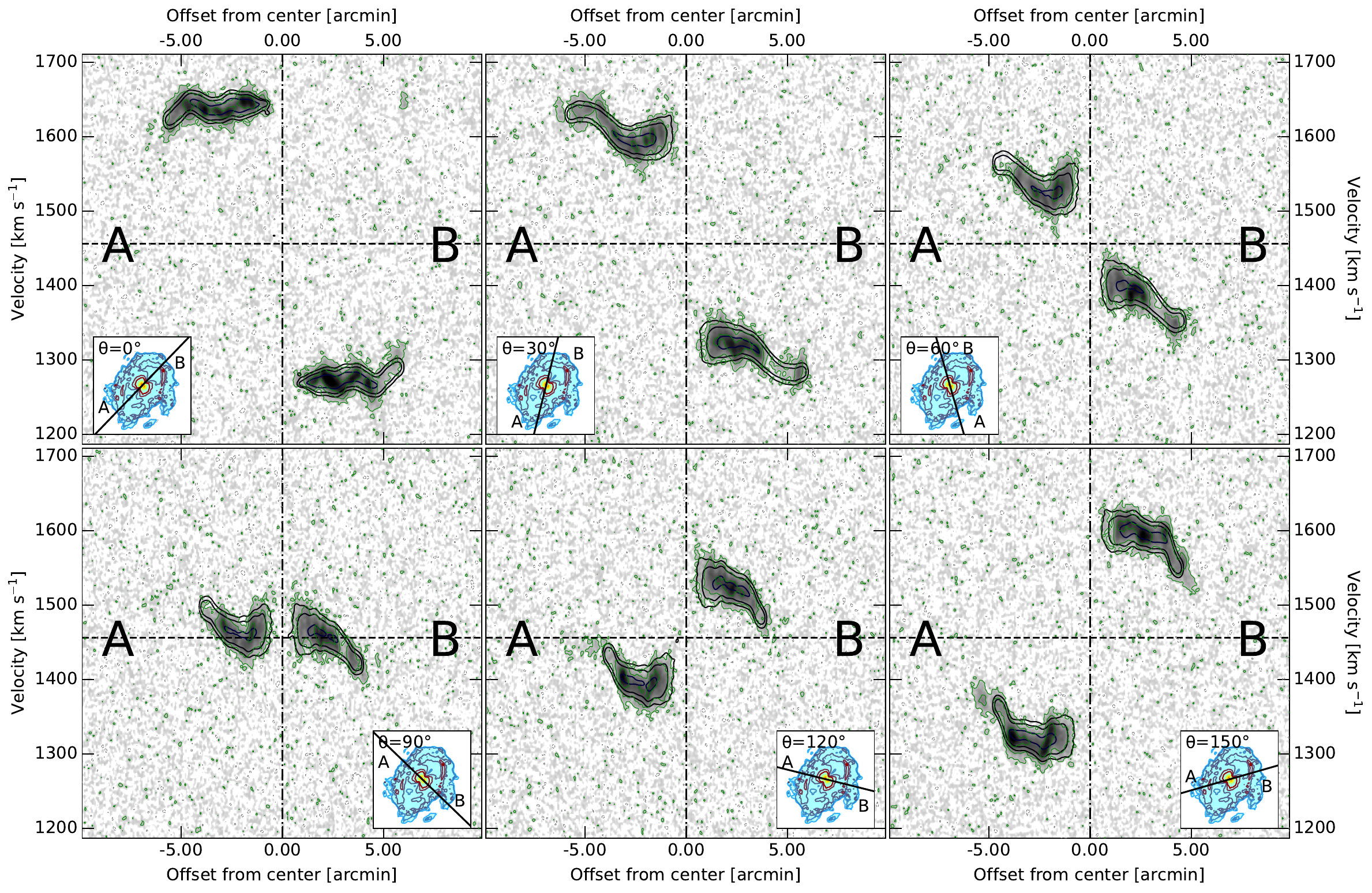}
    \caption{Low-resolution position-velocity diagrams at azimuthal angles ($\theta$) of (0, 30, 60, 90, 120, 150) for the data and the best-fit model derived by \texttt{3D-Barolo} fitting the approaching and receding side of the galaxy simultaneously. Both the data and the model cube were Hanning-smoothed at 7 km s$^{-1}$. For each panel, the background grey-scale image is the slice extracted from the data cube. Solid contour levels are denoting the (4, 16, 64, 256, 1024)$\sigma$. The coloured lines refer to the model, the green curves to the data. The dashed grey contours refers to the $-4\sigma$ level. The dashed black horizontal line corresponds to the systemic velocity, while the dashed-dotted black vertical line denotes the 0-offset. The insets shows the moment 2 map and the slice along which the $pv$ is extracted. Letters $A$ and $B$ provides the orientation in the main panels.}
    \label{fig:pvboth}
\end{figure*}

\begin{figure*}[!h]
    \centering
    \includegraphics[width=.87\hsize]{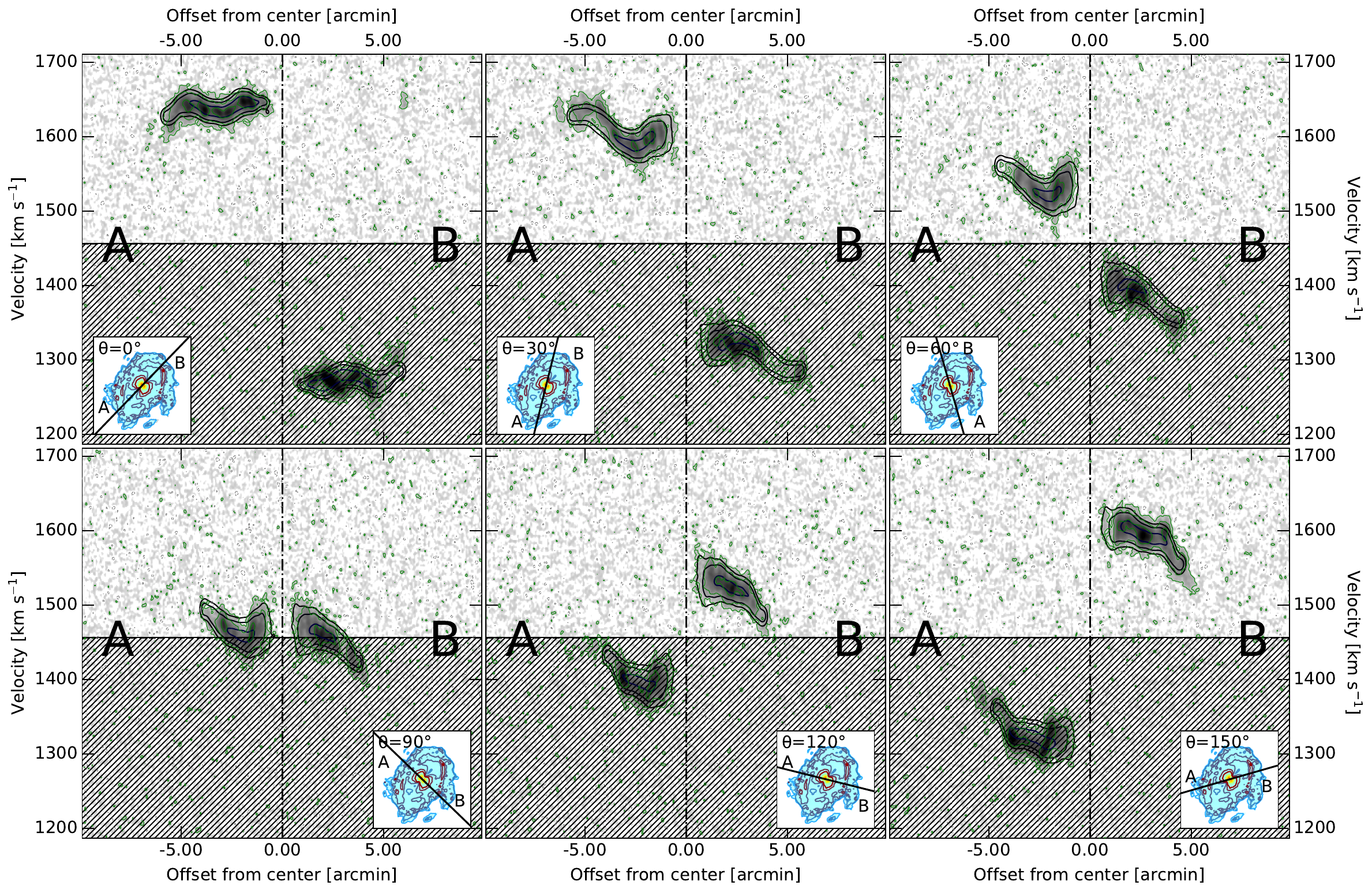}
    \caption{Same as of Fig. \ref{fig:pvboth} but for the fit on the redeing side only. The shaded area indicates the approaching side.}
    \label{fig:pvrec}
\end{figure*}

\begin{figure*}[!h]
    \centering
    \includegraphics[width=.87\hsize]{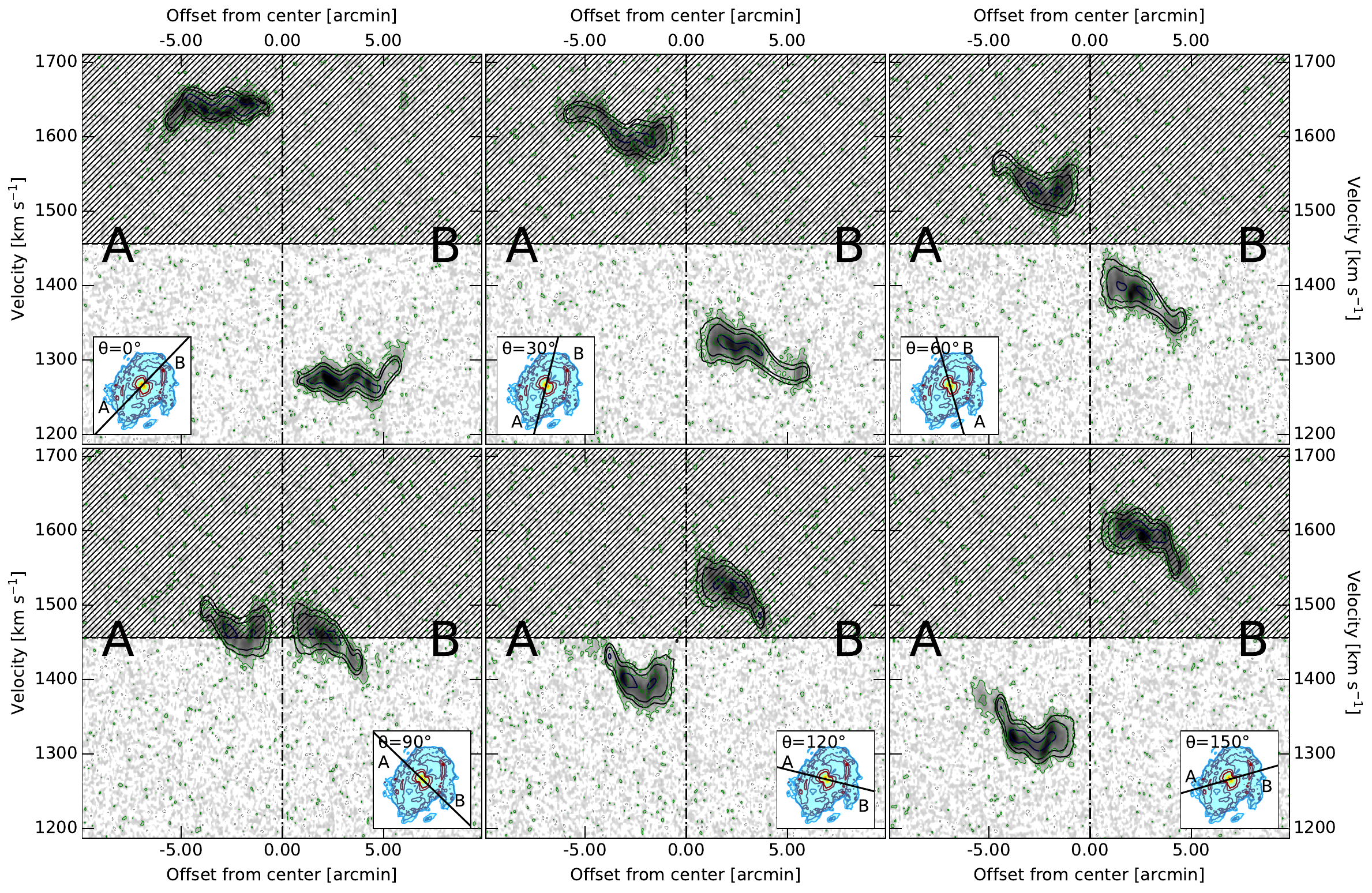}
    \caption{Same as of Fig. \ref{fig:pvboth} but for the fit on the approaching side only. The shaded area indicates the receding side.}
    \label{fig:pvapp}
\end{figure*}

\begin{figure*}
    \centering
    \includegraphics[width=\hsize]{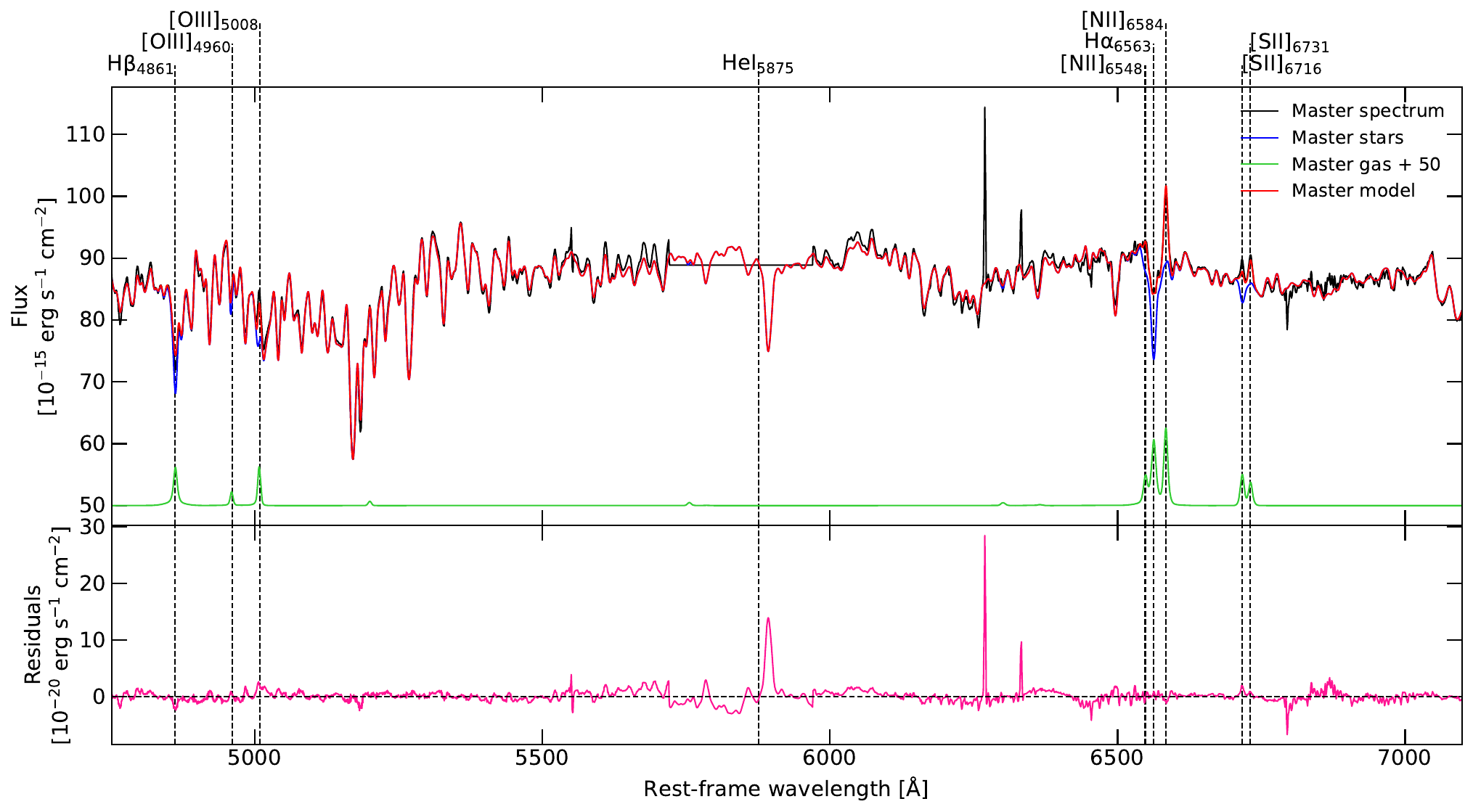}
    \caption{4750-7100\AA{ }integrated spectrum of the inner 5 kpc. The observed spectrum is reported in black, while the best-fit stellar continuum and the emission lines from the ionized gas are shown in blue and green, respectively. The model, i.e., the sum of the blue and green curves, is plotted in red. We shifted the emission lines flux up by $50\times10^{-15}$ erg s$^{-1}$ cm$^{-2}$ for visualization purposes. The bottom panel contains the residuals in terms of data$-$model. The observed 5775-6010\AA{ } range is originally masked because affected by the adaptive optics laser. The flux in these channels is set to the median value of the spectrum prior the fit. The high residuals spikes around 6300\AA{} are not real emission lines but faulty channels.}
    \label{fig:spec}
\end{figure*}

\begin{figure*}
    \centering
    \includegraphics[width=\hsize]{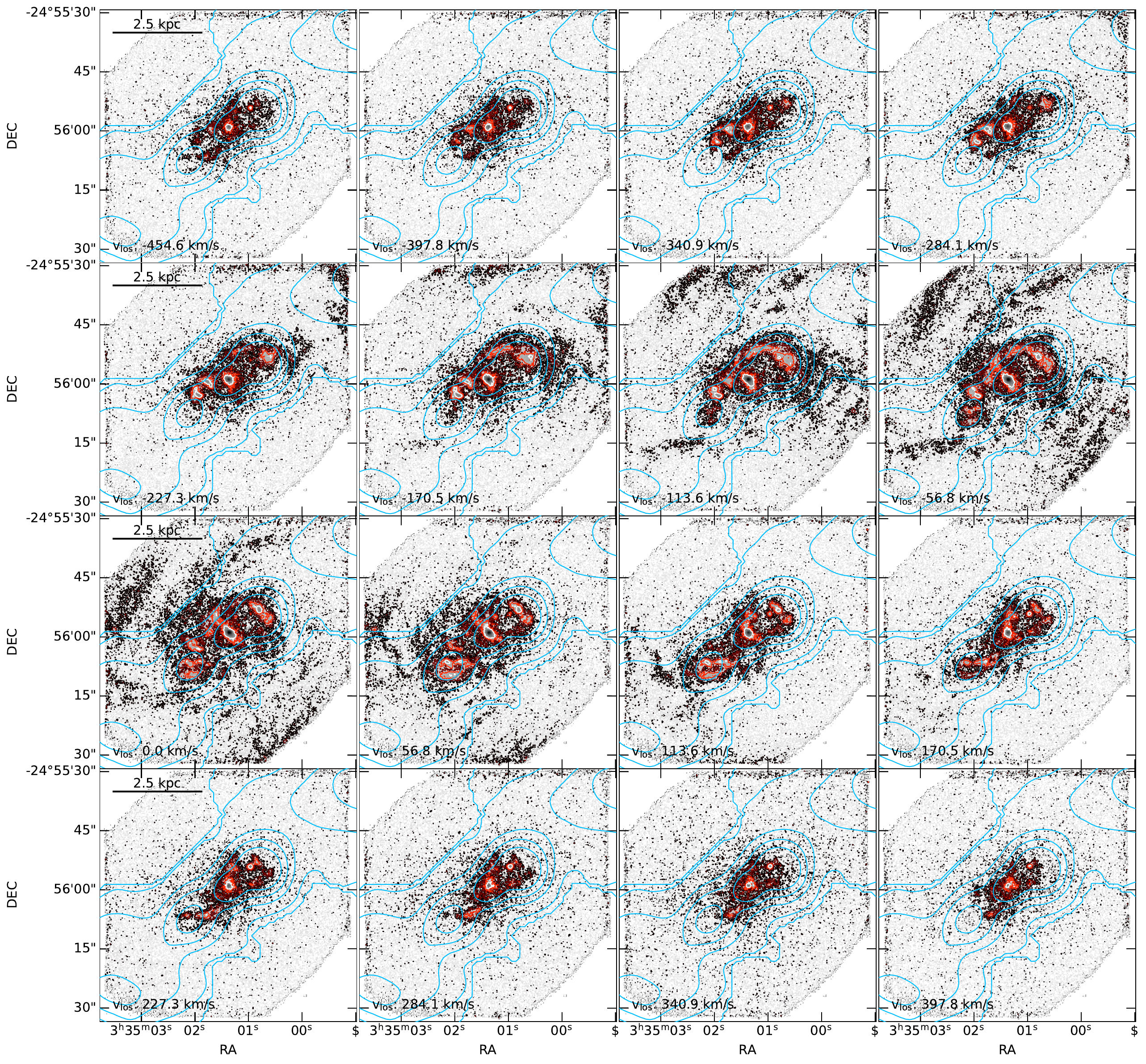}
    \caption{NGC 1371 H$\alpha$ channel maps from $-455$ km s$^{-1}$ to 398 km s$^{-1}$ from the H$\alpha$ systemic velocity in steps of 57 km s$^{-1}$. The colour-scale contours correspond to flux densities of (18, 36, 73, 146) $\times$ 10\textsuperscript{-20} ergs s$^{-1}$ cm$^{-2}$ \AA{}$^{-1}$, equivalent to (4, 8, 16, 32)$\sigma$. The dashed grey contours denote the $-4\sigma$ level. At the bottom, we report the line-of-sight velocity (with respect to the H$\alpha$ systemic velocity) in each channel and at the top left the 2.5-kpc reference scale. The blue contours in all panels indicates the radio continuum (levels: 0.08, 0.31, 0.80, 1.68 mJy beam$^{-1}$)}
    \label{fig:gaschanmap}
\end{figure*}

\begin{figure*}
    \centering
    \includegraphics[width=\hsize]{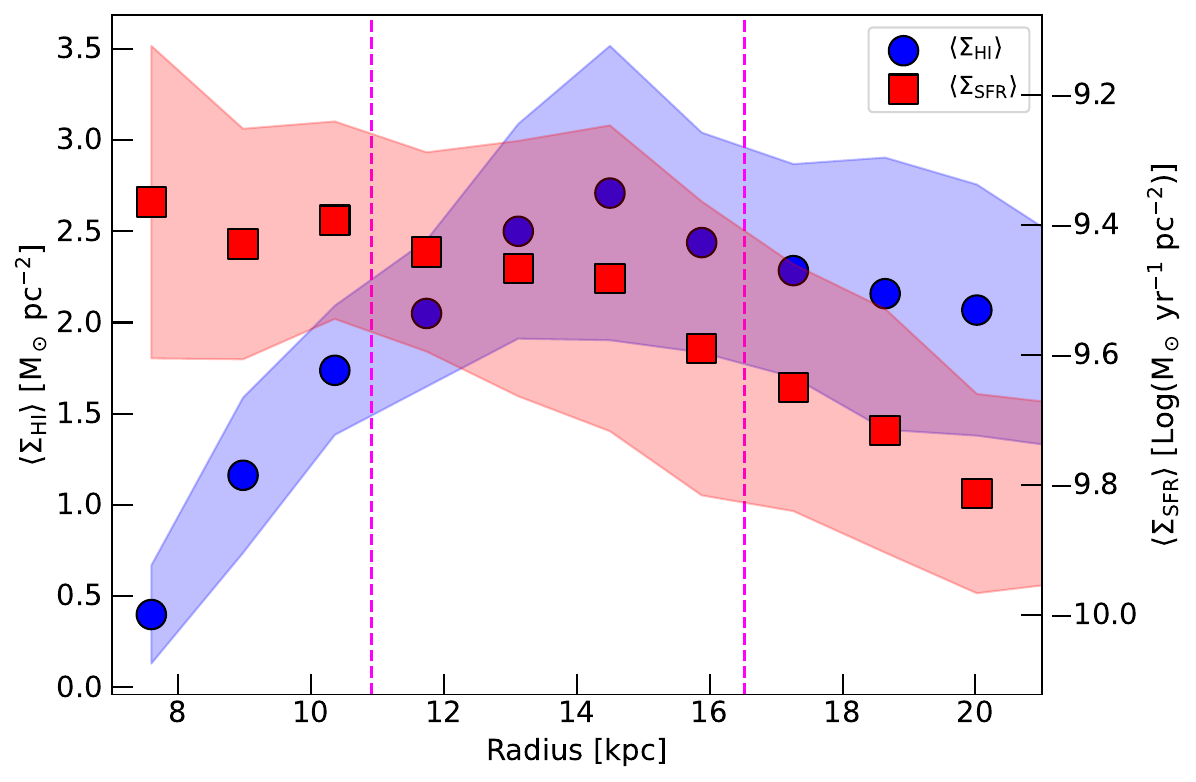}
    \caption{Radial profile of $\Sigma_{\text{SFR}}$ (red) and $\Sigma_{\text{H{\sc i}}}$ (blue). The coloured shaded area is the standard deviation. The vertical magenta dashed lines indicate the approximate location of the star-forming ring. The left y-axis refers to $\Sigma_{\text{H{\sc i}}}$, the right y-axis to $\Sigma_{\text{SFR}}$.}
    \label{fig:sfrhi}
\end{figure*}

\label{LastPage}
\end{document}